\documentclass[aps,prd,superscriptaddress,showpacs,preprint,amsmath,amssymb]{revtex4}
\usepackage{graphicx, bm}
\usepackage[usenames]{color}

\begin{document}

\draft
\title{Sensitivity physics expected to the measurement of the quartic $WW\gamma\gamma$ couplings at the LHeC and the FCC-he}

\author{E. G\"urkanli\footnote{egurkanli@sinop.edu.tr}}
\affiliation{\small Department of Physics, Sinop University, Turkey.\\}

\author{V. Ari\footnote{vari@science.ankara.edu.tr}}
\affiliation{\small Department of Physics, Ankara University, Turkey.\\}

\author{ A. Guti\'errez-Rodr\'{\i}guez\footnote{alexgu@fisica.uaz.edu.mx}}
\affiliation{\small Facultad de F\'{\i}sica, Universidad Aut\'onoma de Zacatecas\\
         Apartado Postal C-580, 98060 Zacatecas, M\'exico.\\}

\affiliation{\small Unidad Acad\'emica de Estudios Nucleares, Universidad Aut\'onoma de Zacatecas,
         98060 Zacatecas, M\'exico.\\}

\author{ M. A. Hern\'andez-Ru\'{\i}z\footnote{mahernan@uaz.edu.mx}}
\affiliation{\small Unidad Acad\'emica de Ciencias Qu\'{\i}micas, Universidad Aut\'onoma de Zacatecas\\
         Apartado Postal C-585, 98060 Zacatecas, M\'exico.\\}

\author{M. K\"oksal\footnote{mkoksal@cumhuriyet.edu.tr}}
\affiliation{\small Deparment of Optical Engineering, Sivas Cumhuriyet University, 58140, Sivas, Turkey.}

\date{\today}

\begin{abstract}

We explore the physics expected sensitivity at the Large Hadron electron Collider (LHeC) and the Future Circular Collider-hadron
electron (FCC-he) to search for the anomalous quartic $WW\gamma\gamma$ couplings in single $W$-boson production in association with a
photon. We study the process $ep \to e^-\gamma^* p \to eW\gamma q'X$ via
the subprocess $\gamma^* q \to W\gamma q'$. The center-of-mass energies and luminosities of the LHeC are assumed to be
$\sqrt{s}=1.30, 1.98$ TeV, ${\cal L}=10-100$ ${\rm fb^{-1} }$ and for the FCC-he $\sqrt{s}=3.46, 5.29$ TeV and
${\cal L}=100-1000$ ${\rm fb^{-1} }$. Considering these energies and luminosities, we estimate sensitivity measures on the anomalous quartic
$WW\gamma\gamma$ couplings at $95\%$ C.L., which can be an order of magnitude more stringent than the experimental limits reported by ATLAS
and CMS Collaborations at the LHC.

\end{abstract}

\pacs{12.60.-i, 14.70.Fm, 4.70.Bh  \\
Keywords: Models beyond the standard model, W bosons, Quartic gauge boson couplings.}

\vspace{5mm}

\maketitle

\section{Introduction}

The $SU(2)_L\times U(1)_Y$ gauge invariant structure of the Standard Model (SM) \cite{SM1,SM2,SM3} specifies the form and
strength of the self-interactions of the vector boson fields, particularly the anomalous Quartic Gauge Couplings (aQGC):
$WW\gamma\gamma$, $WW\gamma Z$, $WWZZ$, $WWWW$, $ZZZZ$, $ZZZ\gamma$, $ZZ\gamma\gamma$, $Z\gamma\gamma\gamma$
and  $\gamma\gamma\gamma\gamma$. Studying which processes these aQGC could contribute to may yield further
confirmation of the non-Abelian gauge structure of the SM or signal the presence of new physics Beyond the SM (BSM) in
unprobed energy scales. For instance, the following present and future colliders: the Large Hadron Collider (LHC), the High-Luminosity Large
Hadron Collider (HL-LHC), the High-Energy Large Hadron Collider (HE-LHC) \cite{HL-HE-LHC}, the Large Hadron electron Collider
(LHeC) \cite{Fernandez,Fernandez1,Fernandez2,LHeC-Bruning}, the Future Circular Collider-hadron electron (FCC-he) \cite{FCChe},
the International Linear Collider (ILC) \cite{ILC-Brau}, the Compact Linear Collider (CLIC) \cite{CLIC-Burrows}, the Circular
Electron Positron Collider (CEPC) \cite{CEPC-Ahmad} and the Future Circular Collider $e^+e^-$ (FCC-ee) \cite{TLEP-Bicer}.
All of these colliders contemplate in their physics programs the study of the aQGC.

Over the last few years, the aQGC production processes and single-$W$ and double-$W$ production in hadron-hadron, lepton-hadron
and lepton-lepton colliders in different collision modes have attracted attention because future colliders
with high energies, high luminosities and cleaner environments may allow experimental studies. Such studies
are interesting because they allow further independent testing of the SM, the quartic $WW\gamma\gamma$
vertex can be probed and the Higgs boson plays an important role in $WW$ channel.

In this paper, we are interested in estimating sensitivity measures in the aQGC $f_{M,i}$ and $f_{T,i}$ with $i = 0, 1, 2,....,7$
for the possible energies and luminosities of the LHeC and the FCC-he in its different stages, i.e., $\sqrt{s}$ = 1.30 TeV,
1.98 TeV, ${\cal L}=$10 fb$^{-1}$, 30 fb$^{-1}$, 50 fb$^{-1}$, 100 fb$^{-1}$ and $\sqrt{s}$ = 3.46 TeV, 5.29 TeV, ${\cal L}=$100
fb$^{-1}$, 300 fb$^{-1}$, 500 fb$^{-1}$, 1000 fb$^{-1}$, respectively. As shown in Figs. 1 and 2, there are 13 Feynman diagrams at the tree level contributing
to the process $ep \to e^-\gamma^* p\to eW\gamma q' X$ via the subprocess $\gamma^* q \to W \gamma q'$, where $q=u, c, \bar d, \bar s$
and $q'=d, s, \bar u, \bar c$.

For an experimental and phenomenological review of the measurement evolution of the limits on the aQGC in the context of previous,
present and future colliders, such as the LEP at the CERN \cite{ALEPH-Barate,DELPHI-Abreu,L3-Acciarri,OPAL-Abbiendi}, D0 and CDF at the Tevatron
\cite{CDF-Gounder,D0-Abbott}, ATLAS and CMS at the LHC \cite{ATLAS-Aaboud,CMS-Chatrchyan} and in the post-LHC era as the LHeC and the FCC-he
\cite{LHeC-FCC-he-WWgg-Ari1,LHeC-FCC-he-WWgg-Ari2}, the ILC, the CLIC, the CEPC and the FCC-ee, see Refs. \cite{Belanger,Stirling,Leil,Bervan,Chong,Koksal,Chen,Stirling1,Aaboud,Atag,Eboli,Eboli1,Sahin,Koksal1,Chapon,Koksal2,Senol,Koksal3,Yang,
Eboli2,Eboli4,Bell,Ahmadov,Schonherr,Wen,Ye,Eboli3,Perez,Sahin1,Senol1,Baldenegro,Fichet,Pierzchala,Gutierrez}, as well as Table I of
Ref. \cite{LHeC-FCC-he-WWgg-Ari2}.

This paper describes searches for the sensitivity physics expected to the measurement of the $WW\gamma\gamma$ aQGC at the LHeC and the
FCC-he using the process $ep \to e^-\gamma^* p \to eW\gamma q'X$. In Section II, we briefly describe the
theoretical aspects of the operators in our effective Lagrangian and in Section III, we derive the bounds for the aQGC
at the LHeC and the FCC-he. We summarize our conclusions in Section IV.

\section{Brief review of the theoretical aspects}

The Effective Field Theory (EFT) approach is very useful in the absence of a specific model of new physics.
An EFT parameterizes the low-energy effects of the new physics to be found at
higher energies in a model-independent way.

With this approach, we start from an EFT to probe model-independent sensitivity measures on $W^+W^-\gamma\gamma$ quartic
gauge boson vertex. The EFT approach is the natural way to extend the SM such that the gauge symmetries are
respected. In addition, the EFT is general enough to capture any BSM physics and provides
guidance as to the most likely place to see the effects of new physics.

The measurement of the $WW\gamma\gamma$ couplings can be made quantitative by introducing a more general
$WW\gamma\gamma$ vertex. For our discussion of phenomenological sensitivities in Section III, we shall use
the phenomenological effective Lagrangian which comes from several $SU(2)_L\times U(1)_Y$ invariant
dimension-8 effective operators \cite{Degrande}:

\begin{equation}
{\cal L}_{eff}=  \sum_{i=1}^2\frac{f_{S, i}}{\Lambda^4}O_{S, i} + \sum_{i=0}^{9}\frac{f_{T, i}}{\Lambda^4}O_{T, i}
+\sum_{i=0}^{7}\frac{f_{M, i}}{\Lambda^4}O_{M, i}.
\end{equation}

\noindent In this equation, the indices S, T and M of the couplings and operators represent three classes of genuine
aQGC operators \cite{Eboli3}. The $f_{T,i}/\Lambda^4$ associated operators
characterize the effect of new physics on the scattering of transversely polarized vector bosons, and $f_{M,i}/\Lambda^4$
includes mixed transverse and longitudinal scatterings. A list of these operators is given below.   \\

i) First class of independent scalar operators:

\begin{eqnarray}
O_{S, 0}&=&[(D_\mu\Phi)^\dagger (D_\nu\Phi)]\times [(D^\mu\Phi)^\dagger (D^\nu\Phi)],  \\
O_{S, 1}&=&[(D_\mu\Phi)^\dagger (D^\mu\Phi)]\times [(D_\nu\Phi)^\dagger (D^\nu\Phi)].
\end{eqnarray}

ii) Second class of independent mixed operators:

\begin{eqnarray}
O_{M, 0}&=&Tr[W_{\mu\nu} W^{\mu\nu}]\times [(D_\beta\Phi)^\dagger (D^\beta\Phi)],  \\
O_{M, 1}&=&Tr[W_{\mu\nu} W^{\nu\beta}]\times [(D_\beta\Phi)^\dagger (D^\mu\Phi)],  \\
O_{M, 2}&=&[B_{\mu\nu} B^{\mu\nu}]\times [(D_\beta\Phi)^\dagger (D^\beta\Phi)],  \\
O_{M, 3}&=&[B_{\mu\nu} B^{\nu\beta}]\times [(D_\beta\Phi)^\dagger (D^\mu\Phi)],  \\
O_{M, 4}&=&[(D_\mu\Phi)^\dagger W_{\beta\nu} (D^\mu\Phi)]\times B^{\beta\nu},  \\
O_{M, 5}&=&[(D_\mu\Phi)^\dagger W_{\beta\nu} (D^\nu\Phi)]\times B^{\beta\mu},  \\
O_{M, 6}&=&[(D_\mu\Phi)^\dagger W_{\beta\nu} W^{\beta\nu} (D^\mu\Phi)],  \\
O_{M, 7}&=&[(D_\mu\Phi)^\dagger W_{\beta\nu} W^{\beta\mu} (D^\nu\Phi)].
\end{eqnarray}

ii) Third class of independent transverse operators:

\begin{eqnarray}
O_{T, 0}&=&Tr[W_{\mu\nu} W^{\mu\nu}]\times Tr[W_{\alpha\beta}W^{\alpha\beta}],  \\
O_{T, 1}&=&Tr[W_{\alpha\nu} W^{\mu\beta}]\times Tr[W_{\mu\beta}W^{\alpha\nu}],  \\
O_{T, 2}&=&Tr[W_{\alpha\mu} W^{\mu\beta}]\times Tr[W_{\beta\nu}W^{\nu\alpha}],  \\
O_{T, 5}&=&Tr[W_{\mu\nu} W^{\mu\nu}]\times B_{\alpha\beta}B^{\alpha\beta},  \\
O_{T, 6}&=&Tr[W_{\alpha\nu} W^{\mu\beta}]\times B_{\mu\beta}B^{\alpha\nu},  \\
O_{T, 7}&=&Tr[W_{\alpha\mu} W^{\mu\beta}]\times B_{\beta\nu}B^{\nu\alpha},  \\
O_{T, 8}&=&B_{\mu\nu} B^{\mu\nu}B_{\alpha\beta}B^{\alpha\beta},  \\
O_{T, 9}&=&B_{\alpha\mu} B^{\mu\beta}B_{\beta\nu}B^{\nu\alpha}.
\end{eqnarray}

\noindent In the operators (2)-(19) $D_\mu$ is the covariant derivative, $\Phi$ denotes the Higgs double field and
$B^{\mu\nu}$, $W^{\mu\nu}$ are the field strength tensors. The  $O_{S, 0}$ and
$O_{S, 1}$ operators given by Eqs. (2) and (3) contain the quartic $W^+W^-W^+W^-$, $W^+W^-ZZ$ and $ZZZZ$ couplings,
which do not concern us here. An exhaustive study on the mechanism to build the dimension-8 operators corresponding to the aQGC is presented
in Refs. \cite{Eboli,Eboli1,Eboli2,Eboli4,Eboli3,Degrande}.

\section{Cross section measurements at the LHeC and the FCC-he}

The phenomenological investigations at $ep$ colliders generally contain usual deep inelastic scattering reactions where the colliding proton
dissociates into partons. These reactions have been extensively examined in the literature, but exclusive and semi-elastic processes that are
$\gamma^* \gamma^*$ and $\gamma^* p$ have been studied much less. These exclusive and semi-elastic processes have simpler final states with
respect to $ep$ processes and thus compensate for the advantages of $ep$ processes such as having a higher center-of-mass energy
and luminosity. Here, $\gamma^* p$ processes have effective luminosity and much higher energy compared to $\gamma^* \gamma^*$ process.
This may be significant because of the high energy dependencies of the cross-sections containing the new physics parameters and for this reason,
$\gamma^* p$ processes are expected to have a high sensitivity to the aQGC.

$\gamma^* p$ processes can be discerned from usual deep inelastic scattering processes by means of two experimental signatures \cite{Rouby}.
The first signature is the forward large rapidity gap \cite{CDF-PRL102-2009,CDF1-PRL102-2009,CMS-JHEP1201-2012,CMS-JHEP1211-2012}.
Quasi-real photons have a low virtuality and scatter with small angles from the beam pipe.
Since the transverse momentum carried by a quasi-real photon is small, photon-emitting electrons should also be scattered with
small angles and exit the central detector without being detected. This causes a decreased energy deposit in the corresponding forward
region. As a result, one of the forward regions of the central detector has a significant lack of energy. This defines the forward
large-rapidity gap, and usual $ep$ deep inelastic processes can be rejected by applying a selection cut on this quantity. The second experimental
signature is provided by the forward detectors \cite{Buniatyan,Li,AFernandez:2012} which are capable of detecting particles with
a large pseudorapidity. When a photon-emitting electron is scattered with a large pseudorapidity, it exceeds the pseudorapidity coverage
of the central detectors. In these processes, the electron can be detected by the forward detectors which provides a distinctive signal for $\gamma^{*}p$ processes.
In this context, LHeC Collaboration has a program of forward physics with extra detectors located in a region between a few tens up to several hundreds of meters
from the interaction point \cite{AFernandez:2012}.

In this section, the cross section of the $ep \to e^-\gamma^* p \to eW\gamma q'X$ signal is evaluated for the center-of-mass energies
and luminosities of the LHeC and the FCC-he with their respective energies and luminosities $\sqrt{s}=1.30, 1.98$ TeV,
${\cal L}=10-100$ ${\rm fb^{-1} }$ and $\sqrt{s}=3.46, 5.29$ TeV, ${\cal L}=100-1000$ ${\rm fb^{-1} }$. For $ep \to e^-\gamma^* p \to eW\gamma q'X$
signal, we consider leptonic and hadronic decays of the $W$-boson; $W \to \nu_l l$, $W \to qq'$ with $\nu_l=\nu_e, \nu_\mu$, $l=e^-, \mu$ and
$q=u, c, \bar d, \bar s$, $q'=d, s, \bar u, \bar c$, respectively.

Formally, the $ep \to e^-\gamma^* p \to eW\gamma q'X$ cross section can be split into three parts:

\begin{eqnarray}
\sigma_{tot}\Biggl( \sqrt{s}, \frac{f_{M,i}}{\Lambda^{4}}, \frac{f_{T,i}}{\Lambda^{4}}\Biggr)&=&\sigma_{BSM}\Biggl(\sqrt{s}, \frac{f^2_{M,i}}{\Lambda^{8}}, \frac{f^2_{T,i}}{\Lambda^{8}},\frac{f_{M,i}}{\Lambda^{4}} \frac{f_{T,i}}{\Lambda^{4}} \Biggr)
+ \sigma_{int}\Biggl( \sqrt{s}, \frac{f_{M,i}}{\Lambda^{4}}, \frac{f_{T,i}}{\Lambda^{4}}\Biggr) \nonumber \\
&+& \sigma_{SM}( \sqrt{s} ), \hspace{5mm}  i=0,...,7,
\end{eqnarray}

\noindent where $\sigma_{BSM}$ is the contribution due to BSM physics, which in our case comes from
the anomalous vertex $WW\gamma\gamma$. $\sigma_{int}$ is the interference term between SM and the new physics
contribution and $\sigma_{SM}$ is the SM prediction, respectively.

To optimize the measurement of the electroweak-induced $eW\gamma q'X$ signal and improve the electroweak
signal significance, we further consider selections on the following variables to suppress backgrounds.
Following is a summary of the baseline selection criteria for the kinematics cuts on the final state particles:  \\

\noindent {\bf i)} {\bf Cuts-0:} Selected cuts for the $p_T$:

\begin{eqnarray}
\bullet  \hspace{3mm}  & p_T^q > 20 \; \mbox{GeV}        &   \mbox{(minimum $p_T$ for the jets)},   \\
\bullet  \hspace{3mm}  & p_T^{\gamma} > 10 \; \mbox{GeV} &   \mbox{(minimum $p_T$ for the photons)},  \\
\bullet  \hspace{3mm}  & p_T^l  >  10\; \mbox{GeV}       &   \mbox{(minimum $p_T$ for the charged leptons)}.
\end{eqnarray}

\noindent {\bf ii)} {\bf Cuts-1:} Selected cuts for the $\eta$:

\begin{eqnarray}
\bullet  \hspace{3mm} & |\eta_q| <  5 &           \mbox{(maximum  rapidity for the jets)},   \\
\bullet  \hspace{3mm} & |\eta_{\gamma}| < 2.5 &    \mbox{(maximum rapidity for the photons)},  \\
\bullet  \hspace{3mm} & |\eta_l| < 2.5 &           \mbox{(maximum rapidity for the charged leptons)}.
\end{eqnarray}

\noindent {\bf iii)} {\bf Cuts-2:} Selected cuts for the $\Delta R$:

\begin{eqnarray}
\bullet  \hspace{3mm} & \Delta R_{qq} = 0.4      &           \mbox{(minimum distance between jets)},    \\
\bullet  \hspace{3mm} &	\Delta R_{ll} = 0.4      &           \mbox{(minimum distance between leptons)},  \\
\bullet  \hspace{3mm} &	\Delta R_{\gamma q} = 0.4&      \mbox{(minimum distance between $\gamma$ and jet)},  \\
\bullet  \hspace{3mm} &	\Delta R_{ql} = 0.4      &           \mbox{(minimum distance between jet and lepton)},  \\
\bullet  \hspace{3mm} &	\Delta R_{\gamma l} = 0.4&     \mbox{(minimum distance between $\gamma$ and lepton)}.
\end{eqnarray}

\noindent As we mentioned above, the kinematic cuts given by Eqs. (21)-(31) are applied to reduce the background and to reach higher
expected significance for the possible aQGC signal in the process $ep \to e^-\gamma^* p \to eW\gamma q'X$. The sensitivities are
investigated using the Monte Carlo simulations with a leading order in MadGraph5\_aMC@NLO \cite{MadGraph}. The operators described
in Eqs. (4)-(19) are implemented into MadGraph5\_aMC@NLO through Feynrules package \cite{AAlloul} as a Universal FeynRules Output (UFO)
module \cite{CDegrande}.

The future lepton-hadron colliders, such as the LHeC and the FCC-he can be operated as $\gamma^*p$ colliders, in this case the emitted
quasi-real photon $\gamma^*$ is scattered with small angles from the beam pipe of $e^{-}$ \cite{Ginzburg,Ginzburg1,Brodsky,Budnev,Terazawa,JYang}.
These processes can be described by the Equivalent Photon Approximation (EPA) \cite{Budnev,Baur1,Piotrzkowski}, using the Weizsacker-Williams
Approximation. The main idea of EPA is that the electromagnetic interaction of an electron with the complicated field of the proton bunch
is replaced by a simpler Compton scattering of this proton with the flux of EPA generated by the electron bunch. For our case, the
schematic diagram for the process $ep \to e^-\gamma^* p \to eW\gamma q'X$ is given by Fig. 1 and the Feynman diagrams of the subprocess
$\gamma^* q \to W\gamma q'$ are shown in Fig. 2. In this context, the spectrum of EPA photons is given by \cite{Budnev,Belyaev}:

\begin{eqnarray}
f_{\gamma^{*}_{1}}(x_{1})=\frac{\alpha}{\pi E_{e}}\{[\frac{1-x_{1}+x_{1}^{2}/2}{x_{1}}]log(\frac{Q_{max}^{2}}{Q_{min}^{2}})-\frac{m_{e}^{2}x_{1}}{Q_{min}^{2}}
&&(1-\frac{Q_{min}^{2}}{Q_{max}^{2}})-\frac{1}{x_{1}}[1-\frac{x_{1}}{2}]^{2}log(\frac{x_{1}^{2}E_{e}^{2}+Q_{max}^{2}}{x_{1}^{2}E_{e}^{2}+Q_{min}^{2}})\} \nonumber \\
\end{eqnarray}

\noindent where $x_{1}=E_{\gamma_{1}^{*}}/E_{e}$ and $Q_{max}^{2}$ is the maximum photon virtuality. The minimum value of $Q_{min}^{2}$ is:

\begin{eqnarray}
Q_{min}^{2}=\frac{m_{e}^{2}x_{1}^{2}}{1-x_{1}}.
\end{eqnarray}

Using all of these tools, the total cross sections (see Eq. (20)) of the $ep \to e^-\gamma^* p \to eW\gamma q'X$ signal at the LHeC
and the FCC-he are determined by:

\begin{eqnarray}
\sigma(ep \to eW\gamma q'X) = \int f_{\gamma^*}(x){\hat\sigma}(\gamma^* q \to W\gamma q') dx.
\end{eqnarray}

The total cross section of the process $ep \rightarrow e^- \gamma^* p \rightarrow eW\gamma q'X $, i.e., $\sigma\left(f_{M,i}/\Lambda^4,
f_{T,i}/\Lambda^4,\sqrt{s}\, \right)$ as a function of $f_{M,i}/\Lambda^4$ and $f_{T,i}/\Lambda^4$ with $i=0,1,2,...,7$ for the energies
of the LHeC with $\sqrt{s}=1.30$ TeV, 1.98 TeV and the FCC-he with $\sqrt{s}=3.46$ TeV, 5.29 TeV are reported in a region defined by the kinematics
cuts given in Eqs. (21)-(31).

Cross sections of the process $ep \rightarrow e^-\gamma^* p \rightarrow eW\gamma q'X$ as a function of aQGC $f_{M,i}/\Lambda^4$ ($f_{T,i}/\Lambda^4$)
are given in Figs. (3)-(10). For evaluation of the total cross sections, the leptonic and hadronic decays of the $W$-boson in the final
state are considered. The total cross sections for each coupling are evaluated while fixing the other couplings to zero.

The corresponding expected cross sections after acceptance cuts for the process $ep \rightarrow e^-\gamma^* p \rightarrow eW\gamma q'X$ give the
value $\sigma\left(f_{T,6}/\Lambda^4,\sqrt{s}\,\right)\simeq 10^5$ pb for $|f_{T,6} /\Lambda^4|=1\times 10^{-8}$ ${\rm GeV^{-4}}$ with the hadronic
decay channel of the $W$-boson. The cross section is $\sigma\left(f_{T,6}/\Lambda^4, \sqrt{s}\, \right)\simeq 10^4$ pb for
the leptonic decay channel of the $W$-boson in the final state.

In Tables I and II, we illustrate the total cross sections in the fiducial region given by Eqs. (21)-(31) for the process
$ep \rightarrow e^-\gamma^* p \rightarrow eW\gamma q'X$ with the different $f_{M,i}/\Lambda^4$ and $f_{T,i}/\Lambda^4$
couplings, and for the future energies of the LHeC and the FCC-he.

From Figs. 3-10 and Tables I and II, it is clear that the cross section projects a greater dependence with respect
to the $f_{T,6}/\Lambda^4$ and $f_{T,5}/\Lambda^4$ couplings than the $f_{M,7}/\Lambda^4$, $f_{M,0}/\Lambda^4$,
$f_{M,1}/\Lambda^4$, etc.. There is also a difference in the measured cross section of up to an order of magnitude between the
leptonic and hadronic cases. The cross sections are evaluated in a region defined by the kinematic cuts
given by Eqs. (21)-(31).

Table III shows the effects of cuts on the cross-section values of SM and some aQGC. As mentioned above, the kinematic cuts
given by Eqs. (21)-(31) are applied to reduce the background and to reach higher expected significance for the possible aQGC
signal in the process $ep \to e^-\gamma^* p \to eW\gamma q'X$. To compare the SM cross section and the total cross sections,
we have taken the values of the couplings as $5\times10^{-10}$ GeV$^{-4}$ for center-of-mass energy of 1.98 TeV and
$5\times10^{-11}$ GeV$^{-4}$ for center-of-mass energy of 5.29 TeV. For example, regarding the effect of cuts at 5.29 TeV
for hadronic decay process, after Cut-0 set is applied, the cut efficiency is about $4\%$ for the SM background which has the
same final state with signal and after applying Cut-1 and Cut-2 sets, the efficiency is reduced by $75\%$. After the
cuts are selected, the SM cross section is more suppressed with respect to the cross sections including the aQGC. Consequently,
sensitivities are better when cuts are applied.

In summary, the $ep \rightarrow e^-\gamma^* p \rightarrow eW\gamma q'X$ cross section in the presence of anomalous couplings increases
rapidly with the electron-proton increasing center-of-mass energy.

\section{$\chi^2$ analysis and sensitivity measures on the aQGC $f_{M,i}/\Lambda^4$ and $f_{T,i}/\Lambda^4$ at the LHeC and the FCC-he}

We perform $\chi^2$ analysis to obtain the sensitivity measures on the anomalous $f_{M,i}/\Lambda^4$ and $f_{T,i}/\Lambda^4$ couplings.
$\chi^2$ is defined as follows:

\begin{equation}
\chi^2(f_{M,i}/\Lambda^4, f_{T,i}/\Lambda^4)=\Biggl(\frac{\sigma_{SM}(\sqrt{s})-\sigma_{BSM}(\sqrt{s}, f_{M,i}/\Lambda^4, f_{T,i}/\Lambda^4)}
{\sigma_{SM}\delta_{st}}\Biggr)^2,
\end{equation}

\noindent where $\sigma_{SM}(\sqrt{s})$ is the cross section of the SM and $\sigma_{BSM}(\sqrt{s}, f_{M,i}/\Lambda^4, f_{T,i}/\Lambda^4)$
is the BSM cross section. $\delta_{st}=\frac{1}{\sqrt{N_{SM}}}$ is the statistical error and $N_{SM}$ is the number of events:

\begin{equation}
N_{SM}={\cal L}_{int}\times \sigma_{SM}.
\end{equation}

\noindent Here, we assume the integrated luminosities ${\cal L}_{int} = 10-100\hspace{0.8mm}{\rm fb^{-1}}$ for the LHeC and
${\cal L}_{int} = 100-1000\hspace{0.8mm}{\rm fb^{-1}}$ for the FCC-he.

Tables IV and V summarize all sensitivity measures on the dimension-8 aQGC obtained from $ep \to e^-\gamma^* p \to eW\gamma q'X$
data with the leptonic decay of the $W$-boson in the final state at center-of-mass energies of $\sqrt{s}$ = 1.30, 1.98 TeV at the LHeC
and $\sqrt{s}$ = 3.46, 5.29 TeV at the FCC-he.
For these sensitivity measures, all parameters except the one shown are fixed to zero. The results for leptonic final state at
$\sqrt{s}$ = 3.46 TeV and $\sqrt{s}$ = 5.29 TeV given in Table V are better values compared to those obtained for the
$\sqrt{s}$ = 1.30 TeV and $\sqrt{s}$ = 1.98 TeV presented in Table IV. A similar behaviour can be seen for the
hadronic decay of $W$-boson given in Tables VI-VII.  The sensitivity measures of $f_{T6}/\Lambda^{4}=[-1.10;1.60]$ ${\rm TeV^{-4}}$
with $\sqrt{s}$ = 3.46 TeV and $f_{T6}/\Lambda^{4}=[-4.37;5.51]\times 10^{-1}$ ${\rm TeV^{-4}}$
with $\sqrt{s}$ = 5.29 TeV in Table VII are the most stringent. These sensitivity measures
are also approximately an order of magnitude more stringent than those obtained at the LHeC and the FCC-he through the main $ep \to e^-\gamma^* p \to eW\gamma q'X $ reaction. \cite{LHeC-FCC-he-WWgg-Ari1,LHeC-FCC-he-WWgg-Ari2}.

Table VIII illustrates sensitivity measures on aQGC at the $95\%$ C. L. via $ep \to e^-\gamma^* p \to eW\gamma q'X $
with the EPA for various $Q_{max}$ values with $\sqrt{s} = 5.29$ TeV at the FCC-he.
The EPA factorize the dependence on virtuality of the photon from the cross-section of the photon-induced process ($\gamma^{*}\gamma^{*}$
and $\gamma^* p$ collisions) to the equivalent photon flux. However, $Q_{max}$ dependence of new physics parameters has also been studied
in the literature. Ref. \cite{Atag-JHEP2010} has examined $Q_{max}$ dependence of the cross sections with the EPA for the process
$pp \to p\gamma^{*}\gamma^{*} p\rightarrow p\tau^{+}\tau^{-}p$ without anomalous couplings of tau lepton.  They found that  the cross sections for $Q_{max}=(1-2)$ ${\rm GeV}$ do not appreciably change. In addition, the cross sections of the process $ee \rightarrow e\gamma^{*} \gamma^{*}e\rightarrow ee\tau\tau$ at the CLIC for values of $Q_{max}=(1.41-8)$ ${\rm GeV}$ are obtained in Ref. \cite{Atag-JHEP2016}. The potential of the process $ep \rightarrow e\gamma^{*}\gamma^{*}p\rightarrow e\tau^{-} \tau^{+} p$ at the LHeC and the FCC-he to examine non-standard $\tau^-\tau^+\gamma$ coupling in a model independent way by means of the effective Lagrangian approach is investigated in Ref. \cite{Koksal-JPG2019}. In that study, $Q_{max}=100$ $\rm GeV$ is assumed and in our case, we consider the maximum photon virtuality $Q_{max}=100$ ${\rm GeV}$, where this value is the default value in MadGraph5\_aMC@NLO. We calculate the $Q_{max}$ dependency on the aQGC for the highest center-of-mass energy. In Table VIII, sensitivity measures on aQGC at the $95\%$ C.L. via the process
$ ep \rightarrow e \gamma^{*} p \rightarrow eW \gamma q^{'}X$ for the hadronic decay of $W$-boson for ${\cal L}=1000$ fb$^{-1}$,
$\sqrt{s}=5.29$ TeV and $Q_{max}=1.41, 8$ ${\rm GeV}$ are obtained. We conclude that $Q_{max}$ dependence on the aQGC does not significantly change.

We now compare our findings with the other results in the literature which used different cuts and different channels.
In Ref. \cite{LHeC-FCC-he-WWgg-Ari1}, a detailed study of the LHeC and the FCC-he sensitivity to the anomalous $f_{M,i}/\Lambda^4$
and $f_{T,i}/\Lambda^4$ couplings in $\nu_e\gamma\gamma q$ production was carried out. Using a $\chi^2$ analysis and kinematic cuts for the final state particles in $\nu_e\gamma\gamma q$ production, they obtained limits on the thirteen different anomalous couplings arising from dimension-8 operators.
In Ref. \cite{LHeC-FCC-he-WWgg-Ari2}, limits were obtained from diboson production at both the LHeC and the FCC-he and
on the anomalous $f_{M,i}/\Lambda^4$ and $f_{T,i}/\Lambda^4$ couplings considering the process $e^-p \to e^-\gamma^*\gamma^*p
\to e^-W^+W^-p$ with the subprocess $\gamma^*\gamma^* \to W^+W^-$. These limits are weaker by about a factor
of 3 or 5 and up to an order of magnitude than our results. CMS Collaboration \cite{CMS-Khachatryan, CMS2-Khachatryan} at the LHC
with  $\sqrt{s} = 8$ TeV and  to an integrated luminosity of 19.7 $\rm fb^{-1}$ searches for exclusive or quasi-exclusive
$WW$ production via the signal topology $pp \to p^*W^+W^-p^* $ where the $p^*$ indicates that the final state protons
either remain intact (exclusive or elastic production), or dissociate into an undetected system (quasi-exclusive
or proton dissociation production). Their research is translated into upper limits on the aQGC operators
$f_{M,0, 1, 2, 3}/\Lambda^4$ (dimension-8). From its investigations, CMS Collaboration
measures the electroweak-induced production of $W$ and two jets, where the $W$ boson decays leptonically,
and experimental limits on aQGC $f_{M,0-7}/\Lambda^4$, $f_{T,0-2, 5-7}/\Lambda^4$ are set at $95\%$ C.L.\cite{CMS-Khachatryan, CMS2-Khachatryan}.
On the other hand, ATLAS Collaboration at the LHC studied the production of $WV\gamma$ events in $e\nu\mu\nu\gamma$,
$e\nu q q \gamma$ and $\mu\nu qq\gamma$ final states with ${\cal L}_{int}=20.2$ ${\rm fb^{-1}}$ of proton-proton collisions
with $\sqrt{s} = 8$ TeV \cite{ATLAS-Aaboud}.

\begin{table}
\caption{Summary of the total cross-sections of the process $ep \to e^-\gamma^* p \to eW\gamma q'X $ for $\sqrt{s}=1.30, 1.98$ TeV
at the LHeC and $\sqrt{s}=3.46, 5.29$ TeV at the FCC-he depending on thirteen anomalous couplings obtained by
dimension-8 operators. The total cross-sections for each coupling are calculated with the values of $1\times10^{-8}$ ${\rm GeV^{-4}}$
and $5\times10^{-9}$ ${\rm GeV^{-4}}$ at the LHeC and the FCC-he, respectively.}
\begin{center}
\begin{tabular}{|c|c|c|c|c|}
\hline\hline
\multicolumn{5}{|c|}{$\sigma(ep \to e^-\gamma^* p \to eW\gamma q'X)$ (pb)}\\
\hline \hline
     & \multicolumn{2}{|c|}{LHeC} & \multicolumn{2}{|c|}{FCC-he}\\
     & \multicolumn{2}{|c|}{Leptonic channel} & \multicolumn{2}{|c|}{Leptonic channel}\\
\hline
SM & 7.27 $\times 10^{-3}$ & 2.18 $\times 10^{-2}$  & 2.27 $\times 10^{-2}$ & 6.38 $\times 10^{-2}$  \\
\hline
Couplings             & $\sqrt{s}= 1.30$ TeV & $\sqrt{s} = 1.98 $ TeV  &  $\sqrt{s} = 3.46 $ TeV & $\sqrt{s} = 5.29 $ TeV  \\
\hline
$f_{M0}/\Lambda^{4}$  & 2.29 $\times 10^{-2}$ & 3.98 $\times 10^{-1}$   & 7.40 $\times 10^{-2}$   &   2.80  \\
\hline
$f_{M1}/\Lambda^{4}$  & 1.66 $\times 10^{-2}$ & 2.29 $ \times 10^{-1}$  & 6.30 $\times 10^{-2}$   &   2.07  \\
\hline
$f_{M2}/\Lambda^{4}$  & 6.79 $\times 10^{-1}$ & 1.62 $\times 10^{1}$    & 2.29                    & 1.17 $\times 10^{2}$ \\
\hline
$f_{M3}/\Lambda^{4}$  & 4.47 $\times 10^{-1}$ & 9.14                    & 1.83                    & 8.61 $\times 10^{1}$ \\
\hline
$f_{M4}/\Lambda^{4}$  & 5.84 $\times 10^{-2}$ & 1.25                    & 1.94 $\times 10^{-1}$   & 9.00 \\
\hline
$f_{M5}/\Lambda^{4}$  & 4.35 $\times 10^{-2}$ & 7.27 $\times 10^{-1}$   & 1.70 $\times 10^{-1}$   & 6.70 \\
\hline
$f_{M7}/\Lambda^{4}$  & 1.05 $\times 10^{-2}$ & 7.69 $\times 10^{-2}$   & 3.29 $\times 10^{-2}$   & 5.71 $\times 10^{-1}$ \\
\hline
$f_{T0}/\Lambda^{4}$  & 9.55 $\times 10^{-1}$ & 3.59 $\times 10^{1}$    & 5.39                    & 5.85 $\times 10^{2}$  \\
\hline
$f_{T1}/\Lambda^{4}$  &  2.61                 & 8.10 $\times 10^{1}$    & 1.96 $\times 10^{1}$    & 1.97 $\times 10^{3}$  \\
\hline
$f_{T2}/\Lambda^{4}$  &  3.25 $\times 10^{-1}$ & 1.02 $\times 10^{1}$   & 2.39                    & 2.36 $\times 10^{2}$  \\
\hline
$f_{T5}/\Lambda^{4}$  & 1.01 $\times 10^{1}$   & 3.88 $\times 10^{2}$   & 5.79 $\times 10^{1}$    & 6.31 $\times 10^{3}$ \\
\hline
$f_{T6}/\Lambda^{4}$  & 2.79 $\times 10^{1}$   & 8.71 $\times 10^{2}$   & 2.12 $\times 10^{2}$    & 2.10 $\times 10^{4}$\\
\hline
$f_{T7}/\Lambda^{4}$  & 3.45                   & 1.10 $\times 10^{2}$   & 26.13                   & 2.51 $\times 10^{3}$\\
\hline
\end{tabular}
\end{center}
\end{table}

\begin{table}
\caption{Summary of the total cross-sections of the process $ep \to e^-\gamma^* p \to eW\gamma q'X $ for $\sqrt{s}=1.30, 1.98$ TeV
at the LHeC and $\sqrt{s}=1.30, 1.98$ TeV at the FCC-he depending on thirteen anomalous couplings obtained by
dimension-8 operators. The total cross-sections for each coupling are calculated with the values of $1\times10^{-8}$ ${\rm GeV^{-4}}$
and $5\times10^{-9}$ ${\rm GeV^{-4}}$ at the LHeC and the FCC-he, respectively.}
\begin{tabular}{|c|c|c|c|c|}
\hline \hline
\multicolumn{5}{|c|}{$\sigma(ep \to e^-\gamma^* p \to eW\gamma q'X )$ (pb)}\\
\hline \hline
     & \multicolumn{2}{|c|}{LHeC} & \multicolumn{2}{|c|}{FCC-he}\\
     & \multicolumn{2}{|c|}{Hadronic channel} & \multicolumn{2}{|c|}{Hadronic channel}\\
\hline
SM & 1.54 $\times 10^{-2}$ & 4.51 $\times 10^{-2}$  & 4.94 $\times 10^{-2}$  & 1.34 $\times 10^{-1}$ \\
\hline
Couplings             & $\sqrt{s} = 1.30$ TeV & $\sqrt{s} = 1.98$ TeV  & $\sqrt{s}=3.46$ TeV   & $\sqrt{s}= 5.29$ TeV \\
\hline \hline
$f_{M0}/\Lambda^{4}$  & 5.79 $\times 10^{-2}$ & 9.38 $\times 10^{-1}$  & 1.79 $\times 10^{-1}$  & 2.50  \\
\hline
$f_{M1}/\Lambda^{4}$  & 6.05 $\times 10^{-2}$ & 6.89 $ \times 10^{-1}$ & 7.34 $\times 10^{-1}$  & 5.88  \\
\hline
$f_{M2}/\Lambda^{4}$  & 1.84          &    3.84 $\times 10^{1}$        & 5.70                   & 1.02 $\times 10^{2}$ \\
\hline
$f_{M3}/\Lambda^{4}$  & 2.08          &    2.82 $\times 10^{1}$        & 2.98 $\times 10^{1}$   & 2.49 $\times 10^{2}$ \\
\hline
$f_{M4}/\Lambda^{4}$  & 1.54 $\times 10^{-1}$  & 2.98                  & 4.79 $\times 10^{-1}$  & 7.88   \\
\hline
$f_{M5}/\Lambda^{4}$  & 1.80 $\times 10^{-1}$  & 2.22                  & 2.32                   & 1.92 $\times 10^{1}$\\
\hline
$f_{M7}/\Lambda^{4}$  & 2.92 $\times 10^{-2}$  & 2.15 $\times 10^{-1}$ & 2.26 $\times 10^{-1}$  & 1.59   \\
\hline
$f_{T0}/\Lambda^{4}$  &  2.27         & 7.24 $\times 10^{1}$           & 1.51 $\times 10^{1}$   & 5.07 $\times 10^{2}$\\
\hline
$f_{T1}/\Lambda^{4}$  &  1.14 $\times 10^{1}$ & 2.15 $\times 10^{2}$   & 5.81 $\times 10^{2}$   & 7.07 $\times 10^{3}$\\
\hline
$f_{T2}/\Lambda^{4}$  &  1.21         & 2.50 $\times 10^{1}$           & 5.11 $\times 10^{1}$   & 6.54 $\times 10^{2}$\\
\hline
$f_{T5}/\Lambda^{4}$  & 2.42 $\times 10^{1}$  & 7.76 $\times 10^{2}$   & 1.62 $\times 10^{2}$   & 5.45 $\times 10^{3}$\\
\hline
$f_{T6}/\Lambda^{4}$  & 1.22 $\times 10^{2}$ & 2.31 $\times 10^{3}$    & 6.28 $\times 10^{3}$   & 7.65 $\times 10^{4}$\\
\hline
$f_{T7}/\Lambda^{4}$  & 1.30 $\times 10^{1}$  & 2.68 $\times 10^{2}$   & 5.50 $\times 10^{2}$   & 7.02 $\times 10^{3}$\\
\hline
\end{tabular}
\end{table}

\begin{table}
\begin{center}
\caption{Effects of selected cuts on the cross-sections of the process $\sigma(ep \to e^-\gamma^* p \to eW\gamma q'X)$
for SM and BSM for various anomalous couplings at 1.98 TeV and 5.29 TeV.
The cross-sections are calculated with the values of $5\times10^{-10}$ ${\rm GeV^{-4}}$
and $5\times10^{-11}$ ${\rm GeV^{-4}}$ at the LHeC and the FCC-he, respectively.
The hadronic decay of $W$-boson is considered.}
\begin{tabular}{|c|c|c|c|c|c|c|c|c|}
\hline \hline
\multicolumn{9}{|c|}{$\sigma(ep \to e^-\gamma^* p \to eW\gamma q'X)$ (pb) }\\
\hline
\hline
   & \multicolumn{4}{|c|}{LHeC}   &     \multicolumn{4}{|c|}{ FCC-he }\\

   & \multicolumn{4}{|c|}{$\sqrt{s}=1.98$ TeV }  & \multicolumn{4}{|c|}{$\sqrt{s}=5.29$ TeV } \\
\hline
Couplings & No Cuts & Cuts-0 & Cuts-1 & Cuts-2  & No Cuts  & Cuts-0  & Cuts-1  & Cuts-2  \\
\hline
  $SM$  & 6.18 & 0.11 & 0.08 & 0.046 &  15.91 & 0.58 & 0.23 & 0.13 \\
\hline
$f_{M0}/\Lambda^{4}$  & 6.42 & 0.12  & 0.08 & 0.05 &  18.16 & 0.57 & 0.25 & 0.14 \\
\hline
$f_{M2}/\Lambda^{4}$  & 7.10 & 0.29  & 0.19 & 0.14 &  19.51 & 1.37  & 0.25 & 0.14 \\
\hline
$f_{M5}/\Lambda^{4}$  & 7.65 & 0.12  & 0.09 & 0.05 &  16.53 & 0.59  & 0.23 & 0.13 \\
\hline
$f_{T0}/\Lambda^{4}$  & 6.91 & 0.53  & 0.30 & 0.22 &  16.68 & 3.65  & 0.44 & 0.17  \\
\hline
$f_{T6}/\Lambda^{4}$  & 12.31 & 8.25 & 7.69 & 5.72 &  80.92 & 65.2  & 33.71 & 7.47 \\
\hline
$f_{T7}/\Lambda^{4}$  & 6.22 & 1.19  & 0.94 & 0.69 &  21.42 & 8.80  & 3.31  & 0.81 \\
\hline
\hline
\end{tabular}
\end{center}
\end{table}

\begin{table}
\caption{Sensitivity measures on aQGC at the $95\%$ C. L. via $ep \to e^-\gamma^* p \to eW\gamma q'X$ for
$\sqrt{s} = 1.30, 1.98$ TeV at the LHeC.}
\begin{tabular}{|c|c|c|c|c|c|}
\hline \hline
\multicolumn{5}{|c|}{$\sqrt{s}$ = 1.30 TeV, \hspace{5mm} {\rm Leptonic channel}}\\
\hline
Couplings (TeV$^{-4}$) & 10 fb$^{-1}$          & 30 fb$^{-1}$                & 50 fb$^{-1}$                   & 100 fb$^{-1}$ \\
\hline
\cline{1-5}
$f_{M0}/\Lambda^{4}$  & [-3.27;3.28] $\times 10^{3}$ & [-2.48;2.49] $\times 10^{3}$ & [-2.18;2.19] $\times 10^{3}$ & [-1.84;1.85] $\times 10^{3}$ \\
\hline
$f_{M1}/\Lambda^{4}$  & [-3.51;4.57] $\times 10^{3}$ & [-2.56;3.62]$ \times 10^{3}$  & [-2.20;3.26] $ \times 10^{3}$ & [-1.78;2.85] $ \times 10^{3}$   \\
\hline
$f_{M2}/\Lambda^{4}$  & [-5.02;5.01] $\times 10^{2}$ & [-3.81;3.80]$ \times 10^{2}$  & [-3.36;3.35] $ \times 10^{2}$ & [-2.82;2.81]$\times 10^{2}$   \\
\hline
$f_{M3}/\Lambda^{4}$  & [-5.39;6.99] $\times 10^{2}$ & [-3.93;5.54]$ \times 10^{2}$  & [-3.38;4.99] $ \times 10^{2}$ & [-2.74;4.35]$\times 10^{2}$   \\
\hline
$f_{M4}/\Lambda^{4}$  & [-1.81;1.82] $\times 10^{3}$ & [-1.37;1.38]$ \times 10^{3}$  & [-1.21;1.22] $ \times 10^{3}$ & [-1.01;1.02]$\times 10^{3}$   \\
\hline
$f_{M5}/\Lambda^{4}$  & [-2.56;1.92] $\times 10^{3}$ & [-2.03;1.39]$ \times 10^{3}$  & [-1.84;1.19] $ \times 10^{3}$ & [-1.61;0.97]$\times 10^{3}$   \\
\hline
$f_{M7}/\Lambda^{4}$  & [-0.91;0.70]$\times 10^{4}$    & [-0.72;0.51]$\times 10^{4}$   & [-0.65;0.44]$\times 10^{4}$   & [-0.57;0.35]$\times 10^{4}$    \\
\hline
$f_{T0}/\Lambda^{4}$  &  [-4.18;4.26] $\times 10^{2}$ & [-3.17;3.24] $\times 10^{2}$ & [-2.78;2.86] $\times 10^{2}$ & [-2.33;2.41] $\times 10^{2}$ \\
\hline
$f_{T1}/\Lambda^{4}$  & [-2.43;2.66] $\times 10^{2}$  & [-1.82;2.05] $\times 10^{2}$ & [-1.59;1.82] $\times 10^{2}$ & [-1.32;1.55] $\times 10^{2}$  \\
\hline
$f_{T2}/\Lambda^{4}$  &  [-0.66;0.79] $\times 10^{3}$ & [-0.49;0.62] $\times 10^{3}$ & [-0.42;0.55] $\times 10^{3}$ & [-0.35;0.48] $\times 10^{3}$ \\
\hline
$f_{T5}/\Lambda^{4}$  & [-1.27;1.30] $\times 10^{2}$  & [-9.64;9.88] $\times 10^{1}$ & [-8.47;8.71] $\times 10^{1}$ & [-7.10;7.35] $\times 10^{1}$  \\
\hline
$f_{T6}/\Lambda^{4}$  & [-7.57;7.94] $\times 10^{1}$  & [-5.71;6.08] $\times 10^{1}$ & [-5.00;5.37] $\times 10^{1}$ & [-4.18;4.55] $\times 10^{1}$ \\
\hline
$f_{T7}/\Lambda^{4}$  & [-2.08;2.34] $\times 10^{2}$  & [-1.55;1.81] $\times 10^{2}$ & [-1.35;1.61] $\times 10^{2}$ & [-1.12;1.38] $\times 10^{2}$ \\
\hline
\multicolumn{5}{|c|}{$\sqrt{s}$ = 1.98 TeV} \\
\hline
\cline{1-5}
$f_{M0}/\Lambda^{4}$  & [-8.86;8.89] $\times 10^{2}$ & [-6.73;6.76] $\times 10^{2}$ & [-5.92;5.95] $\times 10^{2}$ & [-4.97;5.00] $\times 10^{2}$ \\
\hline
$f_{M1}/\Lambda^{4}$  & [-1.07;1.28] $\times 10^{3}$ & [-0.79;1.00]$ \times 10^{3}$  & [-0.68;0.90] $ \times 10^{3}$ & [-0.56;0.78] $ \times 10^{3}$   \\
\hline
$f_{M2}/\Lambda^{4}$  & [-1.35;1.34] $\times 10^{2}$ & [-1.03;1.02]$ \times 10^{2}$  & [-9.00;8.99] $ \times 10^{1}$ & [-7.57;7.56]$\times 10^{1}$   \\
\hline
$f_{M3}/\Lambda^{4}$  & [-1.66;1.92] $\times 10^{2}$ & [-1.24;1.49]$ \times 10^{2}$  & [-1.07;1.33] $ \times 10^{2}$ & [-0.88;1.14]$\times 10^{2}$   \\
\hline
$f_{M4}/\Lambda^{4}$  & [-4.86;4.88] $\times 10^{2}$ & [-3.69;3.70]$ \times 10^{2}$  & [-3.25;3.26] $ \times 10^{2}$ & [-2.73;2.75]$\times 10^{2}$   \\
\hline
$f_{M5}/\Lambda^{4}$  & [-0.71;0.59] $\times 10^{3}$ & [-0.55;0.44]$ \times 10^{3}$  & [-0.50;0.38] $ \times 10^{3}$ & [-0.43;0.31]$\times 10^{3}$   \\
\hline
$f_{M7}/\Lambda^{4}$  & [-2.57;2.14]$\times 10^{3}$    & [-2.01;1.58]$\times 10^{3}$   & [-1.80;1.37]$\times 10^{3}$   & [-1.55;1.12]$\times 10^{3}$    \\
\hline
$f_{T0}/\Lambda^{4}$  &  [-9.00;9.03] $\times 10^{1}$ & [-6.83;6.86] $\times 10^{1}$ & [-6.01;6.04] $\times 10^{1}$ & [-5.05;5.08] $\times 10^{1}$ \\
\hline
$f_{T1}/\Lambda^{4}$  & [-5.91;6.09] $\times 10^{1}$  & [-4.47;4.65] $\times 10^{1}$ & [-3.92;4.10] $\times 10^{1}$ & [-3.28;3.47] $\times 10^{1}$  \\
\hline
$f_{T2}/\Lambda^{4}$  &  [-1.60;1.78] $\times 10^{2}$ & [-1.20;1.38] $\times 10^{2}$ & [-1.04;1.23] $\times 10^{2}$ & [-0.86;1.05] $\times 10^{2}$ \\
\hline
$f_{T5}/\Lambda^{4}$  & [-2.69;2.80] $\times 10^{1}$  & [-2.03;2.14] $\times 10^{1}$ & [-1.78;1.89] $\times 10^{1}$ & [-1.49;1.60] $\times 10^{1}$  \\
\hline
$f_{T6}/\Lambda^{4}$  & [-1.77;1.89] $\times 10^{1}$  & [-1.33;1.45] $\times 10^{1}$ & [-1.17;1.28] $\times 10^{1}$ & [-0.97;1.09] $\times 10^{1}$ \\
\hline
$f_{T7}/\Lambda^{4}$  & [-5.05;5.29] $\times 10^{1}$  & [-3.81;4.05] $\times 10^{1}$ & [-3.34;3.58] $\times 10^{1}$ & [-2.79;3.03] $\times 10^{1}$ \\
\hline \hline
\end{tabular}
\end{table}

\begin{table}
\caption{Sensitivity measures on aQGC at the $95\%$ C. L. via $ep \to e^-\gamma^* p \to eW\gamma q'X $ for
$\sqrt{s} = 3.46, 5.29$ TeV at the FCC-he.}
\begin{tabular}{|c|c|c|c|c|c|}
\hline \hline
\multicolumn{5}{|c|}{$\sqrt{s}$ = 3.46 TeV, \hspace{5mm} {\rm Leptonic channel} }\\
\hline
Couplings (TeV$^{-4}$) & 100 fb$^{-1}$ & 300 fb$^{-1}$ & 500 fb$^{-1}$ & 1000 fb$^{-1}$ \\
\hline \cline{1-5}
$f_{M0}/\Lambda^{4}$  & [-6.70;6.81] $\times 10^{2}$ & [-5.08;5.19] $\times 10^{2}$ & [-4.47;4.57] $\times 10^{2}$ & [-3.75;3.85] $\times 10^{2}$ \\
\hline
$f_{M1}/\Lambda^{4}$  & [-0.62;0.88] $\times 10^{3}$ & [-0.44;0.71]$ \times 10^{3}$  & [-0.38;0.64] $ \times 10^{3}$ & [-0.30;0.57] $ \times 10^{3}$   \\
\hline
$f_{M2}/\Lambda^{4}$  & [-1.03;1.00] $\times 10^{2}$ & [-7.87;7.63]$ \times 10^{1}$  & [-6.95;6.71] $ \times 10^{1}$ & [-5.86;5.62]$\times 10^{1}$   \\
\hline
$f_{M3}/\Lambda^{4}$  & [-1.11;1.13] $\times 10^{2}$ & [-8.42;8.65]$ \times 10^{1}$  & [-7.40;7.63] $ \times 10^{1}$ & [-6.21;6.43]$\times 10^{1}$   \\
\hline
$f_{M4}/\Lambda^{4}$  & [-3.75;3.77] $\times 10^{2}$ & [-2.85;2.87]$ \times 10^{2}$  & [-2.51;2.52] $ \times 10^{2}$ & [-2.11;2.12]$\times 10^{2}$   \\
\hline
$f_{M5}/\Lambda^{4}$  & [-4.17;4.08] $\times 10^{2}$ & [-3.18;3.09]$ \times 10^{2}$  & [-2.80;2.71] $ \times 10^{2}$ & [-2.36;2.27]$\times 10^{2}$   \\
\hline
$f_{M7}/\Lambda^{4}$  & [-1.78;1.39]$\times 10^{3}$    & [-1.40;1.01]$\times 10^{3}$   & [-1.26;0.88]$\times 10^{3}$   & [-1.10;0.71]$\times 10^{3}$    \\
\hline
$f_{T0}/\Lambda^{4}$  &  [-6.71;6.93] $\times 10^{1}$ & [-5.08;5.30] $\times 10^{1}$ & [-4.45;4.67] $\times 10^{1}$ & [-3.73;3.95] $\times 10^{1}$ \\
\hline
$f_{T1}/\Lambda^{4}$  & [-3.49;3.52] $\times 10^{1}$  & [-2.64;2.68] $\times 10^{1}$ & [-2.33;2.36] $\times 10^{1}$ & [-1.95;1.98] $\times 10^{1}$  \\
\hline
$f_{T2}/\Lambda^{4}$  &  [-0.89;1.14] $\times 10^{2}$ & [-0.65;0.90] $\times 10^{2}$ & [-0.56;0.81] $\times 10^{2}$ & [-0.45;0.71] $\times 10^{2}$ \\
\hline
$f_{T5}/\Lambda^{4}$  & [-2.02;2.03] $\times 10^{1}$  & [-1.53;1.55] $\times 10^{1}$ & [-1.35;1.36] $\times 10^{1}$ & [-1.13;1.15] $\times 10^{1}$  \\
\hline
$f_{T6}/\Lambda^{4}$  & [-0.94;1.19] $\times 10^{1}$  & [-0.69;0.94] $\times 10^{1}$ & [-0.60;0.84] $\times 10^{1}$ & [-0.48;0.73] $\times 10^{1}$ \\
\hline
$f_{T7}/\Lambda^{4}$  & [-0.28;0.33] $\times 10^{2}$  & [-0.20;0.26] $\times 10^{2}$ & [-0.18;0.23] $\times 10^{2}$ & [-0.15;0.20] $\times 10^{2}$ \\
\hline
\multicolumn{5}{|c|}{$\sqrt{s}$ = 5.29 TeV}\\
\hline\cline{1-5}
$f_{M0}/\Lambda^{4}$  & [-1.19;1.24] $\times 10^{2}$ & [-9.00;9.45] $\times 10^{1}$ & [-7.90;8.34] $\times 10^{1}$ & [-6.61;7.05] $\times 10^{1}$ \\
\hline
$f_{M1}/\Lambda^{4}$  & [-1.28;1.53] $\times 10^{2}$ & [-0.95;1.20]$ \times 10^{2}$  & [-0.82;1.07] $ \times 10^{2}$ & [-0.67;0.92] $ \times 10^{2}$   \\
\hline
$f_{M2}/\Lambda^{4}$  & [-1.84;1.82] $\times 10^{1}$ & [-1.40;1.38]$ \times 10^{1}$  & [-1.23;1.22] $ \times 10^{1}$ & [-1.04;1.02]$\times 10^{1}$   \\
\hline
$f_{M3}/\Lambda^{4}$  & [-2.06;2.22] $\times 10^{1}$ & [-1.54;1.70]$ \times 10^{1}$  & [-1.35;1.51] $ \times 10^{1}$ & [-1.12;1.28]$\times 10^{1}$   \\
\hline
$f_{M4}/\Lambda^{4}$  & [-6.62;6.68] $\times 10^{1}$ & [-5.02;5.09]$ \times 10^{1}$  & [-4.42;4.48] $ \times 10^{1}$ & [-3.71;3.77]$\times 10^{1}$   \\
\hline
$f_{M5}/\Lambda^{4}$  & [-0.83;0.72] $\times 10^{2}$ & [-0.65;0.54]$ \times 10^{2}$  & [-0.58;0.47] $ \times 10^{2}$ & [-0.49;0.39]$\times 10^{2}$   \\
\hline
$f_{M7}/\Lambda^{4}$  & [-2.98;2.76]$\times 10^{2}$    & [-2.29;2.07]$\times 10^{2}$   & [-2.03;1.81]$\times 10^{2}$   & [-1.73;1.50]$\times 10^{2}$    \\
\hline
$f_{T0}/\Lambda^{4}$  &  [-8.18;8.23] & [-6.21;6.26] & [-5.46;5.51] & [-4.59;4.64] \\
\hline
$f_{T1}/\Lambda^{4}$  & [-4.37;4.60]  & [-3.29;3.52] & [-2.88;3.11] & [-2.41;2.64]  \\
\hline
$f_{T2}/\Lambda^{4}$  &  [-1.23;1.36] $\times 10^{1}$ & [-0.92;1.05] $\times 10^{1}$ & [-0.81;0.93] $\times 10^{1}$ & [-0.67;0.79] $\times 10^{1}$ \\
\hline
$f_{T5}/\Lambda^{4}$  & [-2.41;2.60]  & [-1.81;1.99] & [-1.58;1.77] & [-1.31;1.50]  \\
\hline
$f_{T6}/\Lambda^{4}$  & [-1.22;1.53]  & [-0.89;1.21] & [-0.77;1.08] & [-0.63;0.94] \\
\hline
$f_{T7}/\Lambda^{4}$  & [-3.83;4.07]  & [-2.88;3.13] & [-2.52;2.77] & [-2.10;2.35] \\
\hline
\end{tabular}
\end{table}

\begin{table}
\caption{Sensitivity measures on aQGC at the $95\%$ C. L. via $ep \to e^-\gamma^* p \to eW\gamma q'X $ for
$\sqrt{s} = 1.30, 1.98$ TeV at the LHeC.}
\begin{tabular}{|c|c|c|c|c|c|}
\hline \hline
\multicolumn{5}{|c|}{$\sqrt{s}$ = 1.30 TeV, \hspace{5mm} {\rm Hadronic channel}         }\\
\hline \cline{1-5}
Couplings (TeV$^{-4}$) & 10 fb$^{-1}$ & 30 fb$^{-1}$ & 50 fb$^{-1}$ & 100 fb$^{-1}$ \\
\hline \hline
$f_{M0}/\Lambda^{4}$  & [-2.37;2.41] $\times 10^{3}$ & [-1.80;1.84] $\times 10^{3}$ & [-1.59;1.62] $\times 10^{3}$ & [-1.32;1.36] $\times 10^{3}$ \\
\hline
$f_{M1}/\Lambda^{4}$  & [-1.95;2.59] $\times 10^{3}$ & [-1.41;2.06]$ \times 10^{3}$  & [-1.21;1.86] $ \times 10^{3}$ & [-0.98;1.62] $ \times 10^{3}$   \\
\hline
$f_{M2}/\Lambda^{4}$  & [-3.68;3.61] $\times 10^{2}$ & [-2.81;2.73]$ \times 10^{2}$  & [-2.47;2.40] $ \times 10^{2}$ & [-2.09;2.01]$\times 10^{2}$   \\
\hline
$f_{M3}/\Lambda^{4}$  & [-2.93;3.98] $\times 10^{2}$ & [-2.12;3.17]$ \times 10^{2}$  & [-1.82;2.87] $ \times 10^{2}$ & [-1.47;2.51]$\times 10^{2}$   \\
\hline
$f_{M4}/\Lambda^{4}$  & [-1.31;1.33] $\times 10^{3}$ & [-9.92;10.13]$ \times 10^{2}$  & [-8.72;8.93] $ \times 10^{2}$ & [-7.31;7.53]$\times 10^{2}$   \\
\hline
$f_{M5}/\Lambda^{4}$  & [-1.44;1.07] $\times 10^{3}$ & [-1.15;0.77]$ \times 10^{3}$  & [-1.04;0.66] $ \times 10^{3}$ & [-0.91;0.54]$\times 10^{3}$   \\
\hline
$f_{M7}/\Lambda^{4}$  & [-5.18;3.85]$\times 10^{3}$    & [-4.12;2.79]$\times 10^{3}$   & [-3.73;2.39]$\times 10^{3}$   & [-3.26;1.93]$\times 10^{3}$    \\
\hline
$f_{T0}/\Lambda^{4}$  &  [-3.28;3.29] $\times 10^{2}$ & [-2.49;2.50] $\times 10^{2}$ & [-2.19;2.20] $\times 10^{2}$ & [-1.84;1.85] $\times 10^{2}$ \\
\hline
$f_{T1}/\Lambda^{4}$  & [-1.31;1.63] $\times 10^{2}$  & [-0.96;1.28] $\times 10^{2}$ & [-0.83;1.15] $\times 10^{2}$ & [-0.68;1.00] $\times 10^{2}$  \\
\hline
$f_{T2}/\Lambda^{4}$  &  [-4.06;4.98] $\times 10^{2}$ & [-2.98;3.91] $\times 10^{2}$ & [-2.58;3.50] $\times 10^{2}$ & [-2.11;3.03] $\times 10^{2}$ \\
\hline
$f_{T5}/\Lambda^{4}$  & [-9.54;10.51] $\times 10^{1}$  & [-7.14;8.11] $\times 10^{1}$ & [-6.23;7.20] $\times 10^{1}$ & [-5.17;6.13] $\times 10^{1}$  \\
\hline
$f_{T6}/\Lambda^{4}$  & [-3.85;5.15] $\times 10^{1}$  & [-2.79;4.09] $\times 10^{1}$ & [-2.40;3.70] $\times 10^{1}$ & [-1.94;3.24] $\times 10^{1}$ \\
\hline
$f_{T7}/\Lambda^{4}$  & [-1.23;1.52] $\times 10^{2}$  & [-0.90;1.20] $\times 10^{2}$ & [-0.78;1.08] $\times 10^{2}$ & [-0.64;0.93] $\times 10^{2}$ \\
\hline
\multicolumn{5}{|c|}{$\sqrt{s}$ = 1.98 TeV}\\
\hline \cline{1-5}
$f_{M0}/\Lambda^{4}$  & [-6.76;6.89] $\times 10^{2}$ & [-5.13;5.25] $\times 10^{2}$ & [-4.50;4.63] $\times 10^{2}$ & [-3.78;3.90] $\times 10^{2}$ \\
\hline
$f_{M1}/\Lambda^{4}$  & [-7.10;8.91] $\times 10^{2}$ & [-5.21;7.02]$ \times 10^{2}$  & [-4.49;6.30] $ \times 10^{2}$ & [-3.66;5.47] $ \times 10^{2}$   \\
\hline
$f_{M2}/\Lambda^{4}$  & [-1.05;1.04] $\times 10^{2}$ & [-7.96;7.86]$ \times 10^{1}$  & [-7.02;6.91] $ \times 10^{1}$ & [-5.91;5.80]$\times 10^{1}$   \\
\hline
$f_{M3}/\Lambda^{4}$  & [-1.07;1.38] $\times 10^{2}$ & [-0.78;1.09]$ \times 10^{2}$  & [-0.67;0.98] $ \times 10^{2}$ & [-0.54;0.86]$\times 10^{2}$   \\
\hline
$f_{M4}/\Lambda^{4}$  & [-3.76;3.79] $\times 10^{2}$ & [-2.85;2.89]$ \times 10^{2}$  & [-2.51;2.54] $ \times 10^{2}$ & [-2.10;2.14]$\times 10^{2}$   \\
\hline
$f_{M5}/\Lambda^{4}$  & [-4.90;3.93] $\times 10^{2}$ & [-3.86;2.89]$ \times 10^{2}$  & [-3.46;2.49] $ \times 10^{2}$ & [-3.00;2.03]$\times 10^{2}$   \\
\hline
$f_{M7}/\Lambda^{4}$  & [-1.79;1.42]$\times 10^{3}$    & [-1.41;1.04]$\times 10^{3}$   & [-1.27;0.90]$\times 10^{3}$   & [-1.10;0.73]$\times 10^{3}$    \\
\hline
$f_{T0}/\Lambda^{4}$  &  [-7.54;7.62] $\times 10^{1}$ & [-5.72;5.80] $\times 10^{1}$ & [-5.03;5.11] $\times 10^{1}$ & [-4.23;4.30] $\times 10^{1}$ \\
\hline
$f_{T1}/\Lambda^{4}$  & [-3.99;4.85] $\times 10^{1}$  & [-2.94;3.80] $\times 10^{1}$ & [-2.54;3.40] $\times 10^{1}$ & [-2.08;2.94] $\times 10^{1}$  \\
\hline
$f_{T2}/\Lambda^{4}$  &  [-1.17;1.43] $\times 10^{2}$ & [-0.86;1.12] $\times 10^{2}$ & [-0.74;1.01] $\times 10^{2}$ & [-0.61;0.87] $\times 10^{2}$ \\
\hline
$f_{T5}/\Lambda^{4}$  & [-2.31;2.32] $\times 10^{1}$  & [-1.75;1.76] $\times 10^{1}$ & [-1.54;1.55] $\times 10^{1}$ & [-1.29;1.31] $\times 10^{1}$  \\
\hline
$f_{T6}/\Lambda^{4}$  & [-1.30;1.39] $\times 10^{1}$  & [-0.98;1.07] $\times 10^{1}$ & [-0.86;0.94] $\times 10^{1}$ & [-0.71;0.80] $\times 10^{1}$ \\
\hline
$f_{T7}/\Lambda^{4}$  & [-3.50;4.42] $\times 10^{1}$  & [-2.57;3.49] $\times 10^{1}$ & [-2.21;3.13] $\times 10^{1}$ & [-1.80;2.72] $\times 10^{1}$ \\
\hline
\end{tabular}
\end{table}

\begin{table}
\caption{Sensitivity measures on aQGC at the $95\%$ C. L. via $ep \to e^-\gamma^* p \to eW\gamma q'X $ for
$\sqrt{s} = 3.46, 5.29$ TeV at the FCC-he.}
\begin{tabular}{|c|c|c|c|c|c|}
\hline \hline
\multicolumn{5}{|c|}{$\sqrt{s}$ = 3.46 TeV, \hspace{5mm} {\rm Hadronic channel}  }\\
\hline
Couplings (TeV$^{-4}$) & 100 fb$^{-1}$ & 300 fb$^{-1}$ & 500 fb$^{-1}$ & 1000 fb$^{-1}$ \\
\hline \cline{1-5}
$f_{M0}/\Lambda^{4}$  & [-5.12;5.13] $\times 10^{2}$ & [-3.89;3.90] $\times 10^{2}$ & [-3.42;3.44] $\times 10^{2}$ & [-2.88;2.89] $\times 10^{2}$ \\
\hline
$f_{M1}/\Lambda^{4}$  & [-1.93;2.58] $\times 10^{2}$ & [-1.40;2.05]$ \times 10^{2}$  & [-1.20;1.85] $ \times 10^{2}$ & [-0.97;1.62] $ \times 10^{2}$   \\
\hline
$f_{M2}/\Lambda^{4}$  & [-0.87;0.70] $\times 10^{2}$ & [-0.68;0.52]$ \times 10^{2}$  & [-0.61;0.45] $ \times 10^{2}$ & [-0.53;0.37]$\times 10^{2}$   \\
\hline
$f_{M3}/\Lambda^{4}$  & [-2.94;3.93] $\times 10^{1}$ & [-2.13;3.12]$ \times 10^{1}$  & [-1.83;2.82] $ \times 10^{1}$ & [-1.48;2.47]$\times 10^{1}$   \\
\hline
$f_{M4}/\Lambda^{4}$  & [-2.78;2.89] $\times 10^{2}$ & [-2.10;2.21]$ \times 10^{2}$  & [-1.84;1.95] $ \times 10^{2}$ & [-1.54;1.65]$\times 10^{2}$   \\
\hline
$f_{M5}/\Lambda^{4}$  & [-1.53;1.02] $\times 10^{2}$ & [-1.24;0.73]$ \times 10^{2}$  & [-1.13;0.62] $ \times 10^{2}$ & [-1.00;0.49]$\times 10^{2}$   \\
\hline
$f_{M7}/\Lambda^{4}$  & [-5.29;3.76]$\times 10^{2}$    & [-4.24;2.71]$\times 10^{2}$   & [-3.85;2.32]$\times 10^{2}$   & [-3.39;1.86]$\times 10^{2}$    \\
\hline
$f_{T0}/\Lambda^{4}$  &  [-4.62;5.00] $\times 10^{1}$ & [-3.47;3.85] $\times 10^{1}$ & [-3.03;3.41] $\times 10^{1}$ & [-2.52;2.90] $\times 10^{1}$ \\
\hline
$f_{T1}/\Lambda^{4}$  & [-0.74;0.81] $\times 10^{1}$  & [-0.56;0.62] $\times 10^{1}$ & [-0.49;0.55] $\times 10^{1}$ & [-0.40;0.47] $\times 10^{1}$  \\
\hline
$f_{T2}/\Lambda^{4}$  &  [-2.26;2.99] $\times 10^{1}$ & [-1.64;2.37] $\times 10^{1}$ & [-1.41;2.14] $\times 10^{1}$ & [-1.14;1.87] $\times 10^{1}$ \\
\hline
$f_{T5}/\Lambda^{4}$  & [-1.42;1.51] $\times 10^{1}$  & [-1.07;1.16] $\times 10^{1}$ & [-0.94;1.02] $\times 10^{1}$ & [-0.78;0.87] $\times 10^{1}$  \\
\hline
$f_{T6}/\Lambda^{4}$  & [-2.12;2.62]  & [-1.56;2.05]  & [-1.34;1.84]  & [-1.10;1.60]  \\
\hline
$f_{T7}/\Lambda^{4}$  & [-7.18;8.88]   & [-5.27;6.98]  & [-4.55;6.26]  & [-3.72;5.42]  \\
\hline
\multicolumn{5}{|c|}{$\sqrt{s}$ = 5.29 TeV}\\
\hline \cline{1-5}
$f_{M0}/\Lambda^{4}$  & [-1.53;1.58] $\times 10^{2}$ & [-1.15;1.21] $\times 10^{2}$ & [-1.01;1.07] $\times 10^{2}$ & [-0.85;0.90] $\times 10^{1}$ \\
\hline
$f_{M1}/\Lambda^{4}$  & [-0.90;1.12] $\times 10^{2}$ & [-0.66;0.88]$ \times 10^{2}$  & [-0.57;0.79] $ \times 10^{2}$ & [-0.46;0.68] $ \times 10^{2}$   \\
\hline
$f_{M2}/\Lambda^{4}$  & [-2.41;2.33] $\times 10^{1}$ & [-1.84;1.77]$ \times 10^{1}$  & [-1.62;1.55] $ \times 10^{1}$ & [-1.37;1.30]$\times 10^{1}$   \\
\hline
$f_{M3}/\Lambda^{4}$  & [-1.30;1.84] $\times 10^{1}$ & [-0.94;1.47]$ \times 10^{1}$  & [-0.80;1.34] $ \times 10^{1}$ & [-0.64;1.18]$\times 10^{1}$   \\
\hline
$f_{M4}/\Lambda^{4}$  & [-8.44;8.67] $\times 10^{1}$ & [-6.39;6.62]$ \times 10^{1}$  & [-5.61;5.84] $ \times 10^{1}$ & [-4.70;4.93]$\times 10^{1}$   \\
\hline
$f_{M5}/\Lambda^{4}$  & [-6.23;4.81] $\times 10^{1}$ & [-4.93;3.51]$ \times 10^{1}$  & [-4.44;3.02] $ \times 10^{1}$ & [-3.87;2.45]$\times 10^{1}$   \\
\hline
$f_{M7}/\Lambda^{4}$  & [-2.19;1.83]$\times 10^{2}$    & [-1.71;1.35]$\times 10^{2}$   & [-1.53;1.17]$\times 10^{2}$   & [-1.32;0.96]$\times 10^{2}$    \\
\hline
$f_{T0}/\Lambda^{4}$  &  [-1.04;1.10] $\times 10^{1}$ & [-0.78;0.84] $\times 10^{1}$ & [-0.68;0.74] $\times 10^{1}$ & [-0.57;0.63] $\times 10^{1}$ \\
\hline
$f_{T1}/\Lambda^{4}$  & [-2.53;3.21]   & [-1.85;2.53]  & [-1.60;2.27]  & [-1.30;1.97]   \\
\hline
$f_{T2}/\Lambda^{4}$  &  [-0.82;1.10] $\times 10^{1}$ & [-0.59;0.88] $\times 10^{1}$ & [-0.51;0.79] $\times 10^{1}$ & [-0.41;0.69] $\times 10^{1}$ \\
\hline
$f_{T5}/\Lambda^{4}$  & [-3.17;3.30]  & [-2.39;2.52] & [-2.10;2.23] & [-1.75;1.89]  \\
\hline
$f_{T6}/\Lambda^{4}$  & [-8.17;9.32] $\times 10^{-1}$  & [-6.08;7.23] $\times 10^{-1}$ & [-5.29;6.44] $\times 10^{-1}$ & [-4.37;5.51] $\times 10^{-1}$ \\
\hline
$f_{T7}/\Lambda^{4}$  & [-2.35;3.48]   & [-1.68;2.81]  & [-1.43;2.56]  & [-1.14;2.27]  \\
\hline
\end{tabular}
\end{table}

\begin{table}
\caption{$Q_{max}$ dependence of the sensitivity measures on aQGC at the $95\%$ C. L. via $ep \to e^-\gamma^* p \to eW\gamma q'X $
for ${\cal L}=1000$ fb$^{-1}$ and $\sqrt{s} = 5.29$ TeV at the FCC-he. The hadronic decay of $W$-boson is considered.}
\begin{tabular}{|c|c|c|}\hline
Couplings (TeV$^{-4}$) & $Q_{max}=1.41$ ${\rm GeV}$ & $Q_{max}=8$ ${\rm GeV}$  \\
\hline \hline
$f_{M0}/\Lambda^{4}$  & [-1.09;0.91] $\times 10^{2}$ & [-9.43;9.36] $\times 10^{1}$ \\
\hline
$f_{M1}/\Lambda^{4}$  & [-6.68;6.10] $ \times 10^{1}$ & [-6.81;5.24] $ \times 10^{1}$   \\
\hline
$f_{M2}/\Lambda^{4}$  & [-1.77;1.29] $ \times 10^{1}$ & [-1.59;1.28]$\times 10^{1}$   \\
\hline
$f_{M3}/\Lambda^{4}$  & [-1.17;0.84] $ \times 10^{1}$ & [-1.11;0.79]$\times 10^{1}$   \\
\hline
$f_{M4}/\Lambda^{4}$  & [-5.11;6.12] $ \times 10^{1}$ & [-5.16;5.28]$\times 10^{1}$   \\
\hline
$f_{M5}/\Lambda^{4}$  & [-3.74;3.24] $ \times 10^{1}$ & [-3.84;2.90]$\times 10^{1}$   \\
\hline
$f_{M7}/\Lambda^{4}$  & [-1.14;1.40]$\times 10^{2}$   & [-1.18;1.22]$\times 10^{2}$    \\
\hline
$f_{T0}/\Lambda^{4}$  & [-6.76;6.94] & [-6.12;6.82] \\
\hline
$f_{T1}/\Lambda^{4}$  & [-2.09;1.60] & [-1.89;1.58]  \\
\hline
$f_{T2}/\Lambda^{4}$  & [-6.05;5.98] & [-6.11;5.28] \\
\hline
$f_{T5}/\Lambda^{4}$  & [-3.69;3.72] & [-3.41;3.59]  \\
\hline
$f_{T6}/\Lambda^{4}$  & [-5.07;6.14] $\times 10^{-1}$  & [-4.94;5.60] $\times 10^{-1}$ \\
\hline
$f_{T7}/\Lambda^{4}$  & [-1.75;1.92] & [-1.54;1.93] \\
\hline
\end{tabular}
\end{table}

\section{Conclusions}

The calculations on the production cross section in this paper are derived for the $eW\gamma q'X$ final states at the LHeC
with center-of-mass energies $\sqrt{s}=1.30, 1.98$ TeV and the FCC-he with $\sqrt{s}=3.46, 5.29$ TeV in the fiducial regions
given by Eqs. (21)-(31). Our results are summarized through Figs. 3-10 and in Tables I-III. Furthermore, we show
individual upper sensitivity measures obtained for the aQGC $f_{M,0-5,7}/\Lambda^4$ and $f_{T,0-2,5-7}/\Lambda^4$ at $95\%$ C.L.
both at leptonic and hadronic decay channel of the $W$-boson in Tables IV-VIII. As can be seen in the results, the process gives strong constraints
on aQGC sensitivity measures at high energy region and high luminosities.

In conclusion, we explore the phenomenological aspects of the anomalous $WW\gamma\gamma$ couplings via the process $ep \to e^-\gamma^* p \to eW\gamma q'X$
at the LHeC and the FCC-he. These couplings are defined through a phenomenological effective Lagrangian. The major goal of these measurements will be the confirmation of the new physics BSM. If the energy scale of the new physics responsible for the non-standard gauge boson couplings $f_{M,i}/\Lambda^4$ and $f_{T,i}/\Lambda^4$ is the center-of-mass energy of 5.29 TeV and the integrated luminosity of 1000 ${\rm fb^{-1}}$, these couplings are expected to be no larger than ${\cal O}(10^{-1})$. Our results, as well as our expectations, indicate that with cleaner environments, appropriate fiducial regions, high energies and high luminosities for future colliders will be possible to obtain stronger upper sensitivity measures on the anomalous $ WW\gamma\gamma$ couplings.

\newpage

\begin{center}
{\bf Acknowledgments}
\end{center}

A. G. R. and M. A. H. R. thank SNI and PROFEXCE (M\'exico). The numerical calculations reported in this paper were partially performed at TUBITAK ULAKBIM,
High Performance and Grid Computing Center (TRUBA resources).

\vspace{2cm}


\newpage

\begin{figure}[t]
\centerline{\scalebox{0.65}{\includegraphics{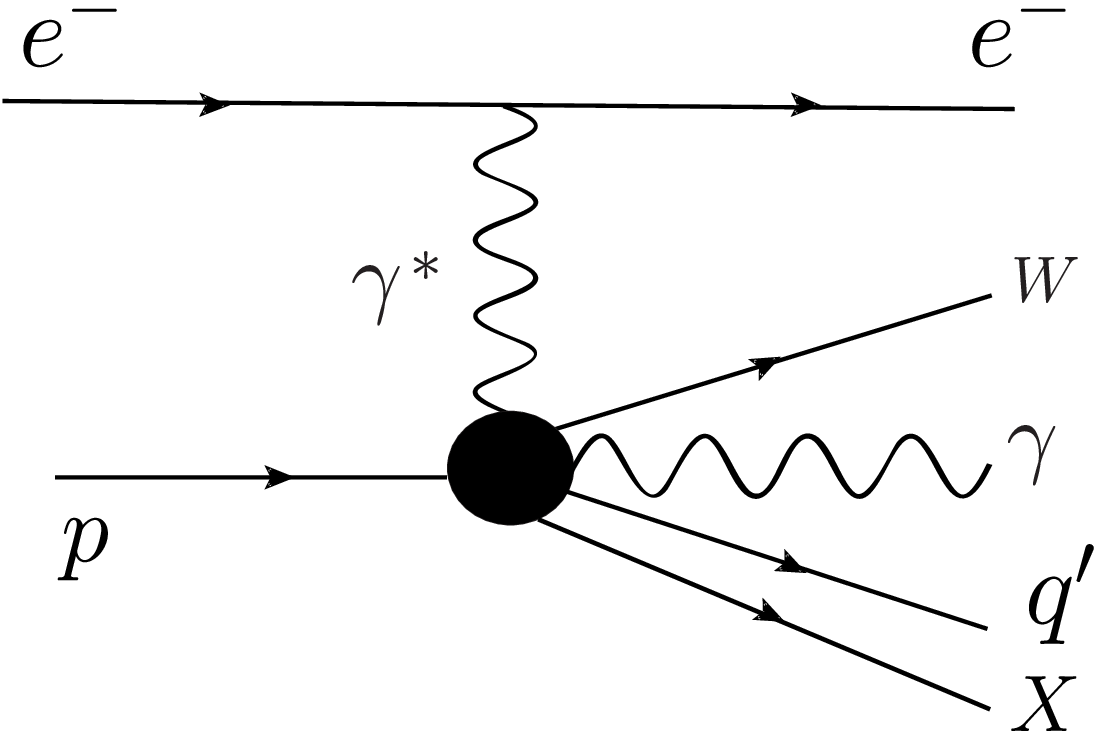}}}
\caption{ \label{fig:gamma1} A schematic diagram for the processes $e^-p \to e^-\gamma^* p \to e^-W \gamma q' X$.}
\label{Fig.1}
\end{figure}

\begin{figure}[t]
\centerline{\scalebox{0.6}{\includegraphics{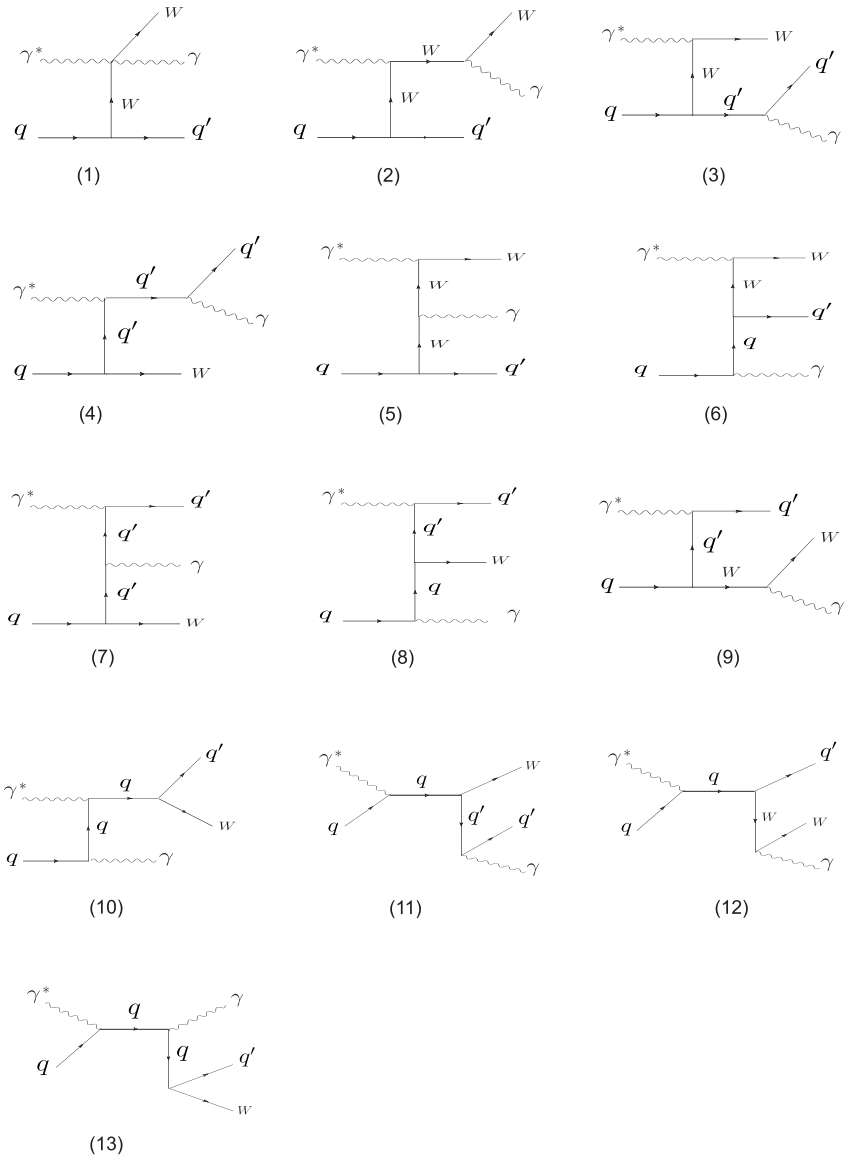}}}
\caption{ \label{fig:gamma2} Feynman diagrams contributing to the subprocess $\gamma^* q \to W\gamma q'$.}
\label{Fig.2}
\end{figure}

\begin{figure}[t]
\centerline{\scalebox{1.3}{\includegraphics{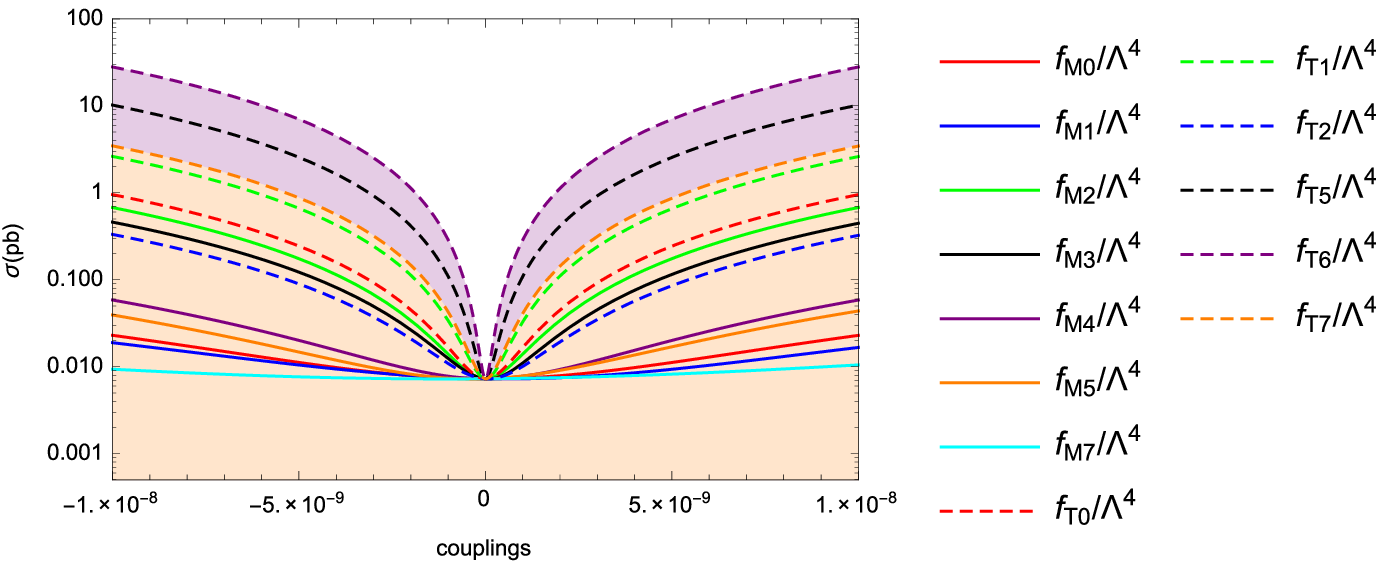}}}
\caption{ \label{fig:gamma1} For pure-leptonic channel, the total cross sections of the process $e^-p \to e^-\gamma^* p \to e^-W^+\gamma q X$
as a function of the anomalous couplings for center-of-mass energy of $\sqrt{s}=1.30$ TeV at the LHeC.}
\label{Fig.1}
\end{figure}

\begin{figure}[t]
\centerline{\scalebox{1.3}{\includegraphics{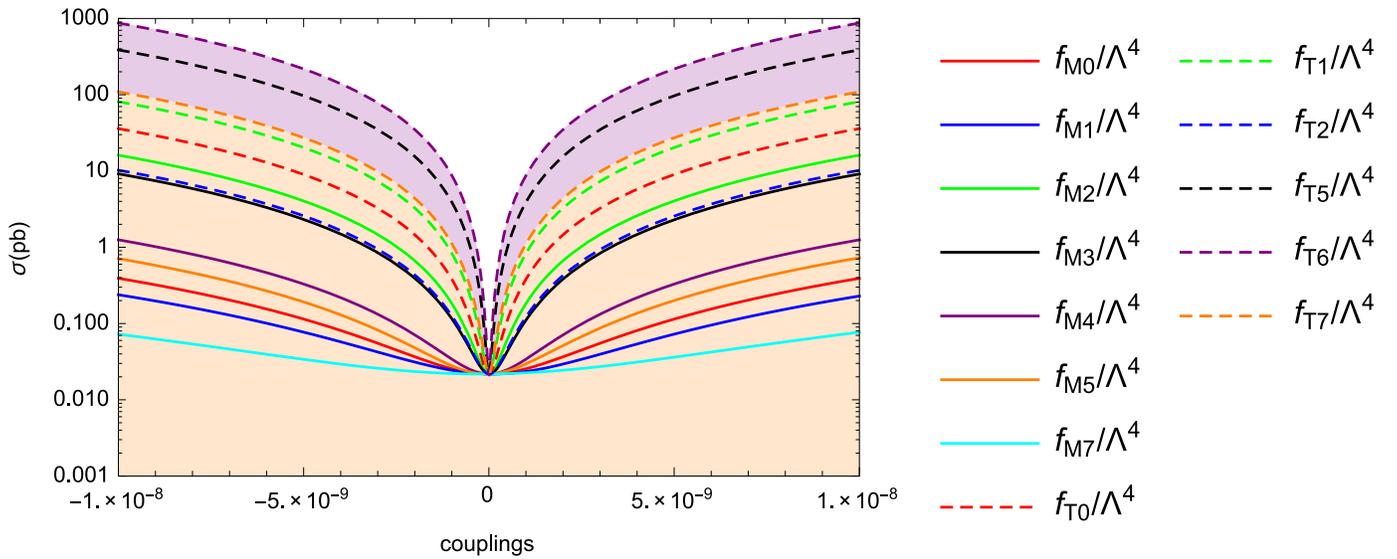}}}
\caption{ \label{fig:gamma2} Same as in Fig. 3, but for $\sqrt{s}=1.98$ TeV at the LHeC.}
\label{Fig.2}
\end{figure}

\begin{figure}[t]
\centerline{\scalebox{1.3}{\includegraphics{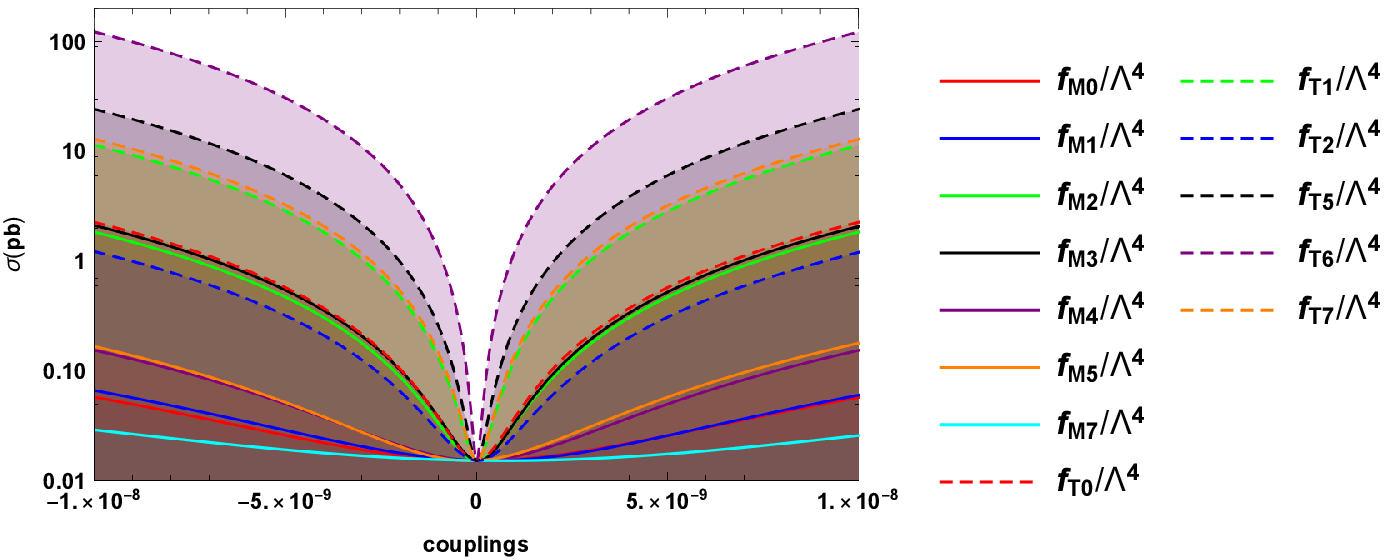}}}
\caption{Same as in Fig. 3, but for hadronic decay.}
\label{Fig.3}
\end{figure}

\begin{figure}[t]
\centerline{\scalebox{1.3}{\includegraphics{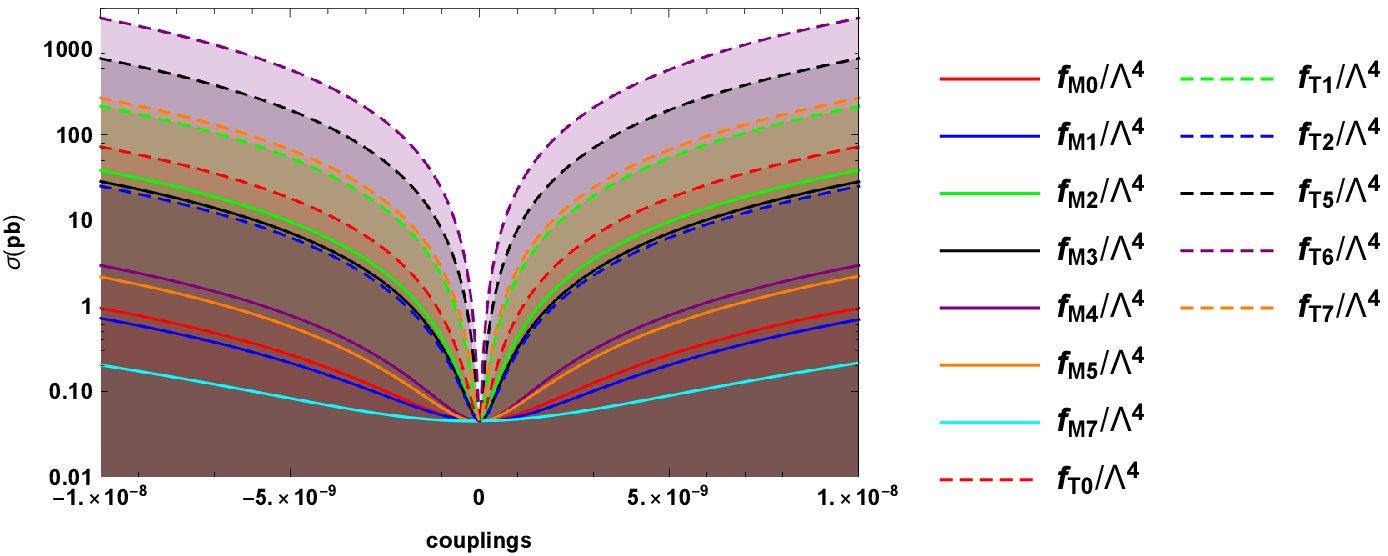}}}
\caption{Same as in Fig. 4, but for hadronic decay.}
\label{Fig.4}
\end{figure}

\begin{figure}[t]
\centerline{\scalebox{1.3}{\includegraphics{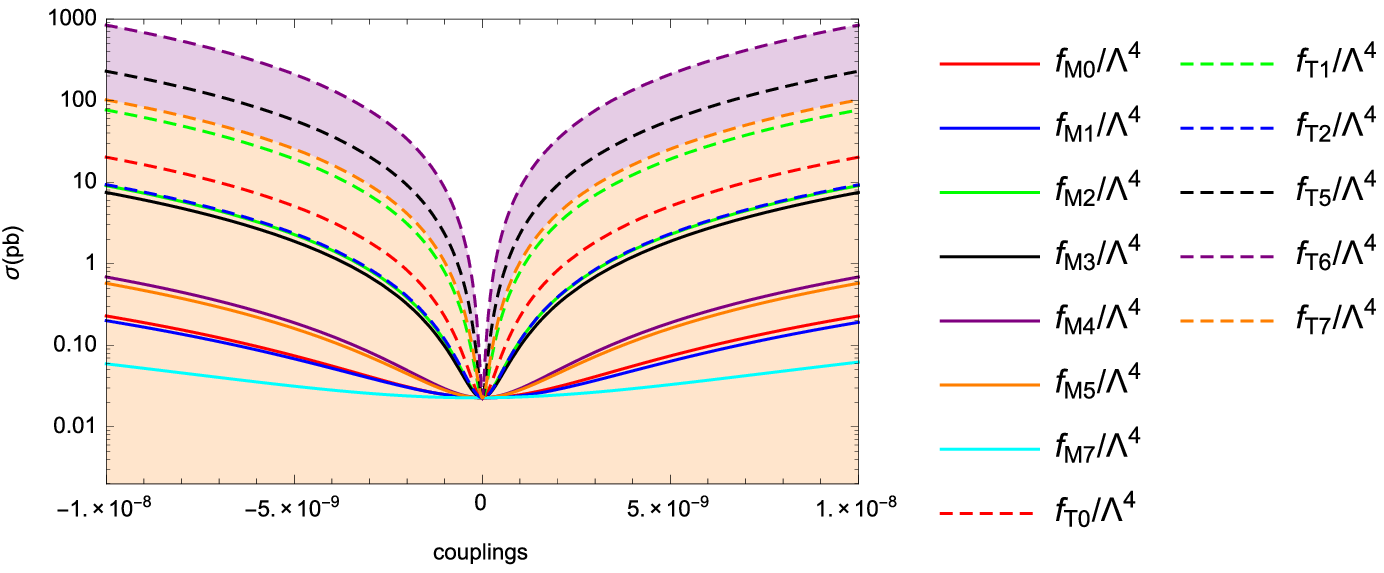}}}
\caption{ \label{fig:gamma1} For pure-leptonic channel, the total cross sections of the process $e^-p \to e^-\gamma^* p \to e^-W^+\gamma q X$
as a function of the anomalous couplings for center-of-mass energy of $\sqrt{s}=3.46$ TeV at the FCC-he.}
\label{Fig.1}
\end{figure}

\begin{figure}[t]
\centerline{\scalebox{1.3}{\includegraphics{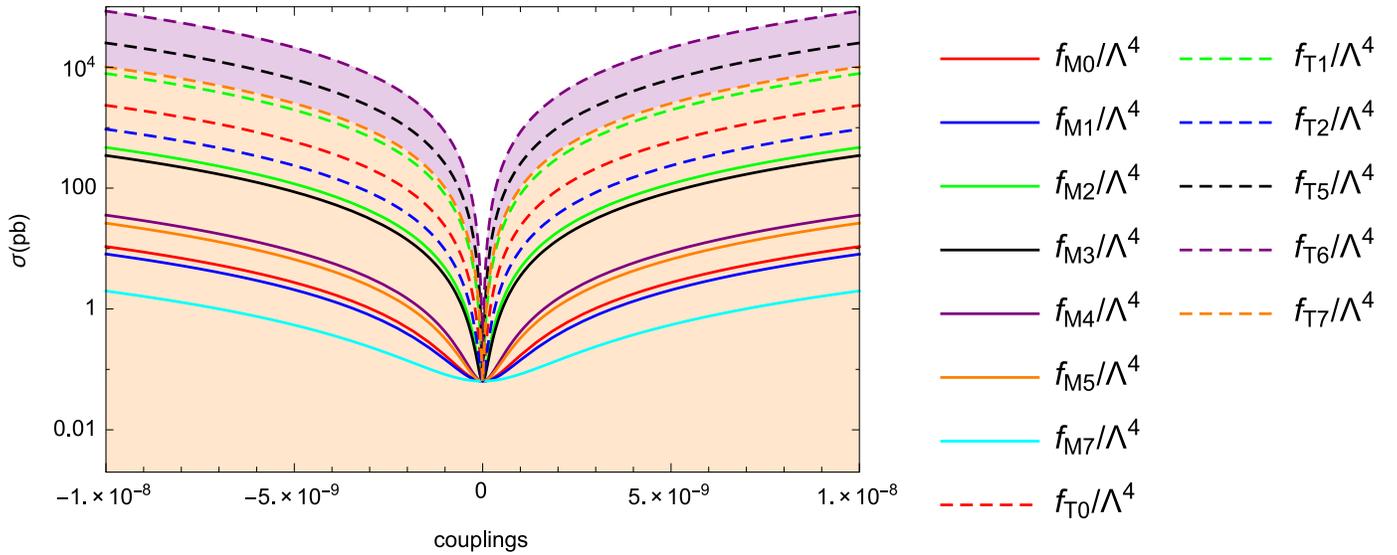}}}
\caption{ \label{fig:gamma2} Same as in Fig. 7, but for $\sqrt{s}=5.29$ TeV at the FCC-he.}
\label{Fig.2}
\end{figure}

\begin{figure}[t]
\centerline{\scalebox{1.3}{\includegraphics{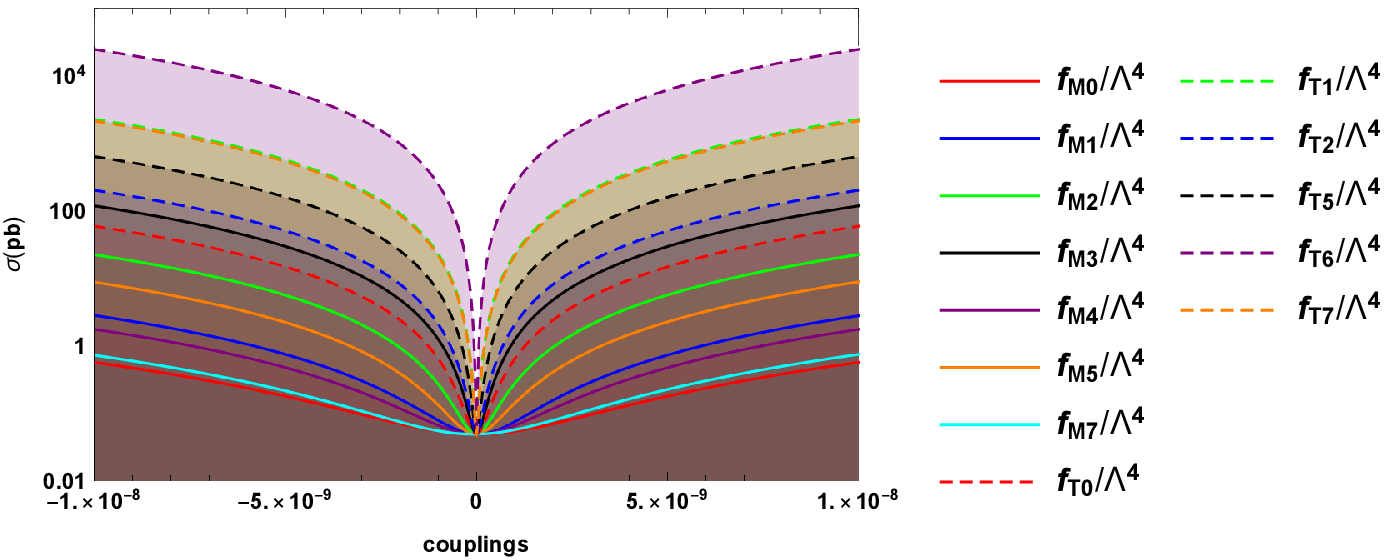}}}
\caption{Same as in Fig. 7, but for hadronic decay.}
\label{Fig.3}
\end{figure}

\begin{figure}[t]
\centerline{\scalebox{1.3}{\includegraphics{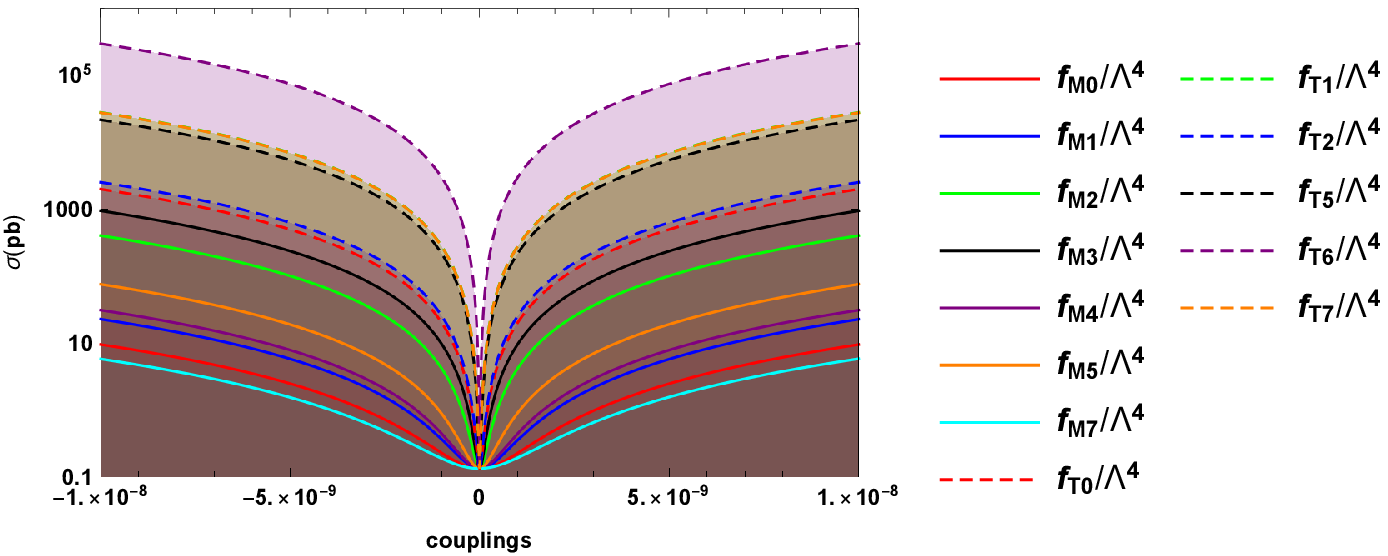}}}
\caption{Same as in Fig. 8, but for hadronic decay.}
\label{Fig.4}
\end{figure}

\end{document}